\documentclass[journal]{IEEEtran}
\usepackage{cite}

\ifCLASSINFOpdf
   \usepackage[pdftex]{graphicx}
\else
\fi
\usepackage{amsmath}
\usepackage{amsfonts}
\usepackage{amssymb,bbm}
\usepackage{mathtools}
\usepackage{booktabs}
\usepackage{float}
\usepackage{adjustbox}
\def\X{{\mathbf X}}
\def\Y{{\mathbf Y}}
\def\V{{\mathbf V}}
\def\P{{\mathbf P}}
\def\Q{{\mathbf Q}}

\def\A{{\mathbf A}}
\def\B{{\mathbf B}}
\def\I{{\mathbf I}}
\def\M{{\mathbf M}}

\def\C{{\mathbf C}}
\def\D{{\mathbf D}}
\def\W{{\mathbf W}}
\def\0{{\mathbf 0}}

\def\K{{\mathbf K}}
\def\U{{\mathbf U}}
\def\J{{\mathbf J}}

\def\x{{\mathbf x}}
\def\l{{\mathbf l}}

\def\u{{\mathbf u}}

\def\z{{\mathbf z}}

\def\R{{\mathbb R}}

\allowdisplaybreaks

\usepackage{algorithmic}

\usepackage{array}
\usepackage{url}

\begin{document}
%
\title{Joint Estimation of Low-Rank Components and Connectivity Graph in High-Dimensional Graph Signals: Application to Brain Imaging}
%
%
%

\author{Rui~Liu,~\IEEEmembership{Student Member,~IEEE,}
        Hossein~Nejati
        and Ngai-Man~Cheung,~\IEEEmembership{Senior Member,~IEEE}
\thanks{Rui Liu, Hossein Nejati and Ngai-Man Cheung are with the Singapore University of Technology and Design (SUTD), Singapore 487372.}
}

\maketitle

\begin{abstract}
This paper presents a graph signal processing algorithm to uncover the intrinsic low-rank components and the  underlying graph of a high-dimensional, graph-smooth and grossly-corrupted  dataset.  
In our problem formulation, we 
assume that the perturbation on the low-rank components is sparse and the signal is smooth on the graph.
We propose an algorithm to  estimate the low-rank components with the help of the graph and refine the graph with better estimated low-rank components.
We propose to perform the low-rank estimation and graph refinement {\em jointly} so that   low-rank estimation can benefit from the refined graph, and  graph refinement can leverage the improved low-rank estimation.  We propose to address the problem with an alternating optimization.
Moreover, we perform a mathematical analysis to understand and quantify the impact of the inexact graph on the low-rank estimation, justifying our scheme with graph refinement as an integrated step in estimating low-rank components.
We perform extensive experiments on the proposed algorithm 
and compare with state-of-the-art low-rank estimation and graph learning techniques.
Our experiments use  synthetic data and real brain imaging (MEG) data that is recorded when subjects are presented with different categories of visual stimuli. We observe that our proposed algorithm is  competitive in estimating the  low-rank components, adequately capturing the intrinsic task-related information in the reduced dimensional representation, 
and leading to better performance in a classification task. Furthermore, we notice that our estimated graph
indicates compatible brain active regions for visual activity as neuroscientific findings.


\end{abstract}


%
\IEEEpeerreviewmaketitle

\section{Introduction}
\label{sec:intro}


\IEEEPARstart{W}{e} consider the problem of uncovering the intrinsic low-rank components of a high-dimensional dataset. 
Furthermore, we focus on data that
resides on a certain graph and the data changes smoothly between
the connected vertices~\cite{shuman2013emerging,ortega:2017}.
We propose to leverage this graph-smoothness property to assist the estimation of low rank components.
However,  in many problems, the underlying graph
is unknown or inexact~\cite{Kalofolias:2016,ortega:2016,petric2017graph}.
In many cases, the graph may be estimated
from the input data which is grossly corrupted. 
Therefore, in this work, we propose to integrate refinement of the graph in the 
low-rank components estimation. As will be discussed, our proposed algorithm estimates the low rank component and refines the underlying graph {\em simultaneously}.


High-dimensional data is common in many engineering areas such as image / video processing, biomedical imaging, computer networks and transportation networks.  As a specific application of our proposed algorithm, we are  interested in automated analysis of brain imaging data. In many cases, the goal is to find the spatiotemporal neural signature of a task, by performing classification on cortical activations evoked by different stimuli \cite{hossein2016, Nejati2015}. 
Common brain imaging techniques are Electroencephalography (EEG) and 
Magnetoencephalography (MEG).  
These measurements are high-dimensional spatio-temporal data.  
Furthermore, the measurements are degraded by various types of noise (e.g., sensor noise, ambient magnetic field noise, etc.) and the noise model is complicated  (non-Gaussian).
The  high-dimensionality and noise  limit both the speed and accuracy of the signal analysis, which may result in unreliable signature modeling for classification.  The high-dimensionality of these signals also increases the complexity of the classifier. 
Combination of a complex classifier and availability of few data samples (due to time, cost, or study limitations) can easily lead to model overfitting. Thus, for a reliable study of brain imaging data, there is a need for a dimensionality reduction method that ensures inclusion of task-related information.

Note that
it has been recognized that there are patterns of anatomical links, statistical dependencies or causal interactions between distinct units within a nervous system~\cite{bullmore2009complex}.  
Some techniques have also been developed to estimate this {\em brain connectivity graph}~\cite{hyde2012cross,brovelli2004beta}.  However, this task is complicated and in many cases the estimated graph may not be accurate.  Thus, it is necessary to have a method to refine the graph.

\subsection{Related Work}

Our work addresses low-rank components estimation and graph estimation as a {\em single integrated scheme}.
Previous work has studied these two problems {\em independently}.
For low-rank components estimation, several linear or nonlinear methods have been proposed that make use of the graph Laplacian of the sample affinity graphs~\cite{Belk03,He:2003,yan:2007,ham2004kernel,bengio2004out,brand2003continuous}.  These methods are geometrically motivated, aiming to preserve the local structures of the data.  The formulations and algorithms of these works are  different from  our work.  
Recently, various approaches have been proposed to incorporate spectral graph regularization~\cite{Jian13b,Shah15a,Shah16b} to uncover the low-rank components.
In all these works,
their graphs are {\em fixed} in their algorithms, precomputed from the noisy input data.
These works do not provision against inaccuracy in the graphs.
On the other hand, as will be discussed, our work studies the impact of the inexact graph on the low-rank components estimation. Our algorithm also refines the graph to improve low-rank components estimation.

Graph estimation has been an active research topic recently~\cite{ortega:2017}.
Different methods have been proposed based on various assumptions on the graph.
The work of Spyrou and Escudero 
\cite{spyrou2017graph} assumes that the original data is a four mode tensor and it applies graph regularized tensor factorization of a four mode tensor to denoise data. 
The work of Maretic et al. \cite{petric2017graph} assumes that the graph signal is sparse and  the signal is a sparse linear combination of a few atoms of a structured graph dictionary. We refer their method as Sparse Graph Dictionary method (\textit{SGDict}) and we compare it with our proposed method in graph estimation accuracy. 
The work of Chepuri et al.  \cite{chepuri2017learning} assumes that the topology is sparse. 
The methods proposed by Pavez and Ortega~\cite{ortega:2016}, Rabbat \cite{rabbat2017inferring} and Hu et al. \cite{lu:2013} assume that the signal is smooth on the graph. In these works, the data, considered as random vectors, is modeled by a Gaussian Markov Random Field (GMRF) distribution, and these works estimate the corresponding precision matrix.  The works of  Kao et al. 
\cite{kao2017disc} and Kalofolias \cite{Kalofolias:2016} combine sparsity and smoothness assumption together and assume that the signal is smooth on the sparse graph. Study on dynamic graph (topology changes along the time) estimation is another active research topic. The work of Villafa{\~n}e-Delgado et al. \cite{villafane2017dynamic} proposes the dynamic Graph Fourier Transform (DGFT) to deal with dynamic graph signal with different Laplacian matrices. The work of Kalofolias et al. \cite{kalofolias2017learning} proposes an algorithm to learn the dynamic graph based on the assumption that the graph is smooth along the time. The work of Shen et al. \cite{shen2017topology} proposes a data model based on autoregressive model to model the time-varying graph signal and learn the graph from the data. 

The work of Dong et al. \cite{Frossard:2016} is most related to our work.
While their focus is to learn the
connectivity graph topology, their algorithm also estimates some
noise-free version of the input data as by-product. Gaussian noise
model and Frobenius norm optimization are employed in \cite{Frossard:2016}. Therefore,
their work is suitable for problem when noise is small. As will be discussed, our 
work assumes a signal model with a low-rank component plus a sparse
perturbation, and an initial graph estimation.
 Our experiment results suggest that our method can perform better with high-dimensional
graph data grossly corrupted by complicated noise, such
as brain imaging signals.
Furthermore, we propose an original analysis on the impact of inexact graph on low-rank components estimation.

Graph signal processing tools have been applied to some brain imaging tasks.
In \cite{huang2016graph}, graph Fourier transform is applied to decompose brain signals into low, medium, and high frequency components for analysis of functional brain networks properties. The work of Liu et al. 
\cite{rui2016dimensionality} uses the eigenvectors of the graph Laplacian to approximate the intrinsic subspace of high-dimensional brain imaging data.  They evaluate different brain connectivity estimations to compute the graph.
Estimation of the brain connectivity graph using a Gaussian noise model has been proposed in \cite{lu:2013}. The work of Griffa et al. \cite{griffa2017transient} proposes a method to captures the network of spatio-temporal connectivity integrating neural oscillations and anatomical connectivity from diffusion fMRI. The work of Medaglia et al. \cite{medaglia2017functional} applies graph Fourier transform to distill functional brain signals into aligned and liberal components. 
The work of Guo et al. \cite{guo:2017} applies deep convolutional networks on graph signals for brain signal analysis.
Graph signal processing has also been shown to be useful in  image compression \cite{hu2015multiresolution}, temperature data \cite{frossard2015}, wireless sensor data \cite{egilmez2014spectral} and Internet traffic analysis \cite{manas:2017}.  A few signal features motivated by graph signal processing have also been proposed~\cite{dong2013inference, kang2014complex}. 

\subsection{Our Contribution}

In this paper, we propose a robust algorithm, {\em low-rank components and graph estimation (LGE)}, to estimate the low-rank components and refine the graph simultaneously. Based on a signal model of sparse perturbation and graph smoothness,  the proposed LGE estimates the low-rank components of the data using the graph smoothness assumption and iteratively refines the graph with the estimated low-rank components.  Our low-rank components estimation includes graph refinement as an integrated step with the aim 
to improve the effectiveness of the graph smoothness constraint during 
low-rank components estimation, 
 increasing the quality of estimated low-rank components.
Our specific contributions are:
\begin{itemize}
\item Based on a model with a low-rank component plus a sparse perturbation, and an initial graph estimation, we propose an algorithm to simultaneously learn the low-rank component and the graph.
\item We derive the learning steps using alternating optimization and ADMM.
\item We conduct an analysis on  the impact of inexact graph on the  low-rank estimation accuracy. 
\item We perform extensive experiments using synthetic and real brain imaging data in a supervised classification task.
\end{itemize}
This work is an extension of our previous conference paper \cite{rui2017simultaneous}. Compared with \cite{rui2017simultaneous}, we make several improvements of the algorithm. Moreover, we conduct a detailed analysis of the impact of graph distortion on the low-rank component estimation.  Also, we conduct thorough experiments and compare with recently proposed methods.

The rest of the paper is organized as follows. We present our model, objective function, and learning algorithm in Section~\ref{sec:lowrank_GraphEstimate}. Then, we describe a detailed analysis on the impact of graph distortion in Section~\ref{sec:SensitivityAnalysis}. We provide detailed experiment results using synthetic data in Section~\ref{sec:Synthetic}. We present experiment results using real brain imaging data in Section~\ref{sec:BrainData}. Section~\ref{sec:conclusion} concludes the work.


\section{Simultaneous low-rank components and graph estimation}
\label{sec:lowrank_GraphEstimate}

\begin{table*}\small 
	\caption{A summary of notations used in this work}
	\label{tab:Symbol_list}
	\begin{tabular}{c|c}
		\toprule[1pt]
		\textbf{Notation}&\textbf{Terminology}\\
		\hline
		$p$ & Number of features/nodes in the graph\\
		\hline
		$n$ & Number of samples/instances \\
		\hline
		$k$ & Index of iteration \\
		\hline
		$\gamma$, $\delta$, $\beta$ & Parameter used in LGE \\
		\hline
		$r$ & Rank of low-rank components matrix\\
		\hline
		$r_1$, $r_2$  & Step size used in ADMM \\
		\hline
		$\tau_1 = 2/(r_1+r_2)$, $\tau_2 = \delta/r_1$  & Threshold values in the first step of LGE \\
		\hline
		$g$, $h= \tau/ave(\sum_{i=1}^p\sigma_i\cdot \mathbbm{1}(\sigma_i\leq \tau))$ & Approximation parameters \\
		\hline
		$q$ & Probability of connected nodes \\
		\hline
		$d$ &  Distortion probability of perturbation matrix\\
		\hline
		$s$ & Distortion probability of graph Laplacian matrix \\
		\hline
		$u$, $c$ & Distortion probability and amplitude of low-rank components matrix \\
		\hline
		$A$, $B$ &  Two brain regions\\
		\hline
		$E_a(t)$, $E_b(t)$ &  Amplitudes of the signals for region $A$ and $B$ at time $t$\\
		\hline
		$\psi_a(t)$, $\psi_b(t)$ &  Phrases of the signals for region $A$ and $B$ at time $t$ \\
		\hline
		$\X \in \R^{p \times n}$ & Input data matrix\\
		\hline
		$\P \in \R^{p \times r}$ & Principal components \\
		\hline
		$\Q \in \R^{n \times r}$ & Projected data points \\
		\hline
		$\Y \in \R^{n \times r}$ & Coefficient matrix \\
		\hline
		$\textbf{L} \in \R^{p \times n}$, $\tilde{\l_i} \in \R^n$  & Low-rank components of data matrix with $\tilde{\l_i}$ as the $i$-th row vector of $\textbf{L}$\\
		\hline
		$\textbf{L}_0 \in \R^{p \times n}$ & Ground truth low-rank components matrix of data matrix\\
		\hline
		$\hat{\textbf{L}} \in \R^{p \times n}$ & Estimated low-rank components matrix\\
		\hline
		$\textbf{L}^{(k)} \in \R^{p \times n}$ & Estimated low-rank components matrix in the $k-th$ iteration\\
		\hline
		$\tilde{\textbf{L}} = \textbf{L}_0 + \Delta\textbf{L} \in \R^{p \times n}$ & Distorted low-rank components matrix with $\Delta \textbf{L}$ as the distortion \\
		\hline
		$\M \in \R^{p \times n}$ & Perturbation matrix\\
		\hline
		$\mathcal{G} =(\mathcal{V},\mathcal{E}, \W)$ & Graph with $\mathcal{V}$ as a set of vertices, $\mathcal{E}$ as a set of edges and $\W \in \R^{p \times p}$ as adjacency matrix \\
		\hline
		$\D \in \R^{p \times p}$ & Diagonal degree matrix of $\W$ \\
		\hline
		$\Phi_f \in \R^{p \times p}$ & Laplacian matrix of graph $\mathcal{G}$\\
		\hline
		$\Phi_{f0} \in \R^{p \times p}$ & Ground truth Laplacian matrix of graph $\mathcal{G}$\\
		\hline 
		$\hat{\Phi_f} \in \R^{p \times p}$ & Estimated Laplacian matrix of graph $\mathcal{G}$\\
		\hline  
		$\tilde{\Phi_f} = \Phi_{f0} + \Delta\Phi_f \in \R^{p \times p}$ & Distorted Laplacian matrix of graph $\mathcal{G}$ with $\Delta \Phi$ as the distortion \\
		\hline	
		$\z_1 \in \R^{p \times n}$, $\z_2 \in \R^{p \times n}$ & Lagrange multipliers used in ADMM\\
		\hline
		$\K \in \R^{p \times n}$, $\J= \U\Sigma\V^T \in \R^{p \times n}$ & Intermediate matrices calculated in ADMM\\
		\hline
		$\U \in \R^{p \times p}$, $\V \in \R^{n \times p}$ & Left and right singular vectors of $\J$\\
		\hline
		$\Sigma = diag(\sigma_i) \in \R^{p \times p}$ & The rectangular diagonal matrix with singular values $\sigma_i$ of $\J$ on the diagonal \\
		\hline	
		$D_{\tau}(\J) \triangleq \U\Omega_{\tau}(\Sigma)\V^T$ & Singular value decomposition (SVD)  of matrix $\J$ with soft-threshold value $\tau$\\
		\hline
		$\Omega_{\tau}(x) \triangleq sgn(x)max(|x|-\tau,0)$ & Element-wise soft-threshold operator\\
		\hline
		\bottomrule[1pt]		
	\end{tabular}
	\centering
	\vspace{+0.1cm}
	\normalsize
\end{table*}

Table \ref{tab:Symbol_list} lists the important notations used in this work.
We consider   $\X=(\x_1,\dots,\x_n) \in \R^{p \times n}$,
the high-dimensional data matrix that consists of $n$ $p$-dimensional data points. 
For our brain imaging data, 
$\X$ are the measurements by the $p$ sensors at the $n$ time instants, i.e., $p$ time series.
We assume that the data points have low intrinsic dimensionality and lie near some low-dimensional subspace and propose the following mathematical model for the data: 
\begin{equation}
\X = \textbf{L}_0 + \M
\end{equation}
$\textbf{L}_0 \in \R^{p \times n}$ is the low-rank components of the data matrix which is of primary interest in this paper, and $\M  \in \R^{p \times n}$ is a perturbation matrix that has arbitrarily large magnitude with sparse support. 

Principal component analysis (PCA) is the most popular technique for determining the low-rank component with applications including image, video, signal, web content and network. 
The \emph{classic PCA} finds the projection $\Q^T \in \R^{r \times n}$ of $\X$ in a $r$-dimensional ($r \le p$) linear space characterized by an orthogonal basis $\P \in \R^{p \times r}$, by solving the following optimization:
\begin{equation}
\begin{aligned}
& \underset{\P,\Q}{\text{minimize}}
& & \| \X-\P\Q^T\|_F^2 \\
& \text{subject to}
& & \P^T \P = \I
\end{aligned}
\end{equation}
The $\P$ and $\Q^T$ matrices are known as principal components and projected data points, respectively. $\textbf{L} = \P\Q^T \in \R^{p \times n}$ is the approximation of the
low-rank component.
The classic PCA suffers from a few disadvantages.  
First, 
it is susceptible to grossly corrupted data in $\X$.
Second, it does not consider the implicit data manifold information.

Candes \emph{et al.} \cite{Cand11} addressed the first issue by designing \emph{Robust PCA (RPCA)}, which is robust to outliers by directly recovering the low-rank matrix $\textbf{L}$ from the grossly corrupted $\X$:
\begin{equation}
\begin{aligned}
& \underset{\textbf{L},\M}{\text{minimize}}
& & \|\textbf{L}\|_* + \delta \|\M\|_1 \\
& \text{subject to}
& & \X = \textbf{L} + \M
\end{aligned}
\label{eq:rpca}
\end{equation}
$\|.\|_*$ denotes the nuclear norm which 
is used as a convex surrogate of rank. $\M$ is the corresponding perturbation matrix. 

In this work we propose to extend (\ref{eq:rpca}) with an additional graph smoothness regularization, while the underlying graph topology that captures the data correlation could be  {\em unknown} or {\em inexact} (thus some refinement is needed):
\begin{equation}
\begin{aligned}
& \underset{\textbf{L},\M, \Phi_f}{\text{minimize}}
& & \|\textbf{L}\|_* + \delta \|\M\|_1 +   \gamma tr(\textbf{L}^T \Phi_f \textbf{L}) + \beta \|\Phi_f \|_F^2\\
& \text{subject to}
& & \X = \textbf{L} + \M, \\
&&& \Phi_f \in \mathcal{L}
\end{aligned}
\label{eq:rpca-ug}
\end{equation}
$\|.\|_F$ denotes the Frobenius norm. Here $\Phi_f$ is the graph Laplacian matrix of the feature graph  $\mathcal{G}$ describing the correlation between individual features: $\mathcal{G} = (\mathcal{V},\mathcal{E}, \W)$ consists of a finite set of vertices $\mathcal{V}$ ($|\mathcal{V}| = p$), a set of edges  $\mathcal{E}$, and a weighted adjacency matrix $\W = \{ W_{i,j} | W_{i,j} \ge 0 \}$ ($W_{i,j}$ quantifying the similarity between the $i$-th and $j$-th features of the $p$-dimensional measurement vectors). Graph Laplacian matrix is defined as $\Phi_f = \D-\W$, with 
$\D$ being the diagonal degree matrix.
$\mathcal{L}$ is the set of all valid $p \times p$ graph Laplacian matrix $\Phi$:
\begin{equation}
\mathcal{L}= \{ \Phi : \Phi_{ij} = \Phi_{ji} \leq 0, \Phi_{ii} = - \sum_{j \neq i} \Phi_{ij}  \}
\end{equation}

We solve (\ref{eq:rpca-ug}) iteratively using alternating minimization with the following justifications:
\begin{itemize}
\item
{\bf $\textbf{L}, \M$ given $\Phi_f$:} 
For a given $\Phi_f$ (even a rough estimate),   $tr(\textbf{L}^T \Phi_f \textbf{L})$  imposes an additional constraint on the underlying (unknown) low-rank components $\textbf{L}$.  Specifically, 
\begin{equation}
tr(\textbf{L}^T \Phi_f \textbf{L}) = \frac{1}{2} \sum_{i,j}  W_{i,j} \|    \tilde{\l_i} -  \tilde{\l_j}   \|^2,
\end{equation}
where $\tilde{\l_i} \in \R^n$ is the $i$-th row vector of $\textbf{L}$.
Therefore, $tr(\textbf{L}^T \Phi_f \textbf{L})$ in (\ref{eq:rpca-ug}) forces the row $i$ and $j$ of $\textbf{L}$ to have similar values if $W_{i,j}$ is large.  Note that in our brain imaging data, individual rows represent the time series captured by sensors. Thus, $tr(\textbf{L}^T \Phi_f \textbf{L})$ forces the low-rank components of the time series to be similar for highly correlated sensors.  Prior information regarding measurement correlation (such as the physical distance between the capturing sensors) can be incorporated as the initial $\Phi_f$  to bootstrap the estimation of $\textbf{L}$. 
\item
{\bf $\Phi_f$ given $\textbf{L}, \M$:}
For a given estimation of the low-rank components $\textbf{L}$, $tr(\textbf{L}^T \Phi_f \textbf{L})$ guides the refinement of $\Phi_f$ and hence the underlying connectivity graph $\mathcal{G}$. In particular, a graph $\mathcal{G}$ that is consistent with the signal variation
in $\textbf{L}$ is favored: large $W_{i,j}$ if row $i$ and $j$ of $\textbf{L}$ have similar values.  In many problems, the given graph for a problem can be noisy (e.g., the graph is estimated from the noisy data itself \cite{Shah15a,Shah16b}).  The proposed formulation iteratively improves $\Phi_f$ using the refined low-rank components.  The improved $\Phi_f$
 in turn facilitates the low-rank components estimation. 
\end{itemize}

In this work, 
we propose to solve equation (\ref{eq:rpca-ug}) with an alternating minimization scheme where, at each step, we fix one or two variables and update the other variables.
Specifically,  
in the first step, for a given $\Phi_f$, we solve the following optimization problem using alternating direction method of multipliers (ADMM)\cite{boyd2011distributed} with respect to $\textbf{L}$ and $\M$:
\begin{equation}
\begin{aligned}
& \underset{\textbf{L},\M}{\text{minimize}}
& & \|\textbf{L}\|_* + \delta \|\M\|_1 +   \gamma tr(\textbf{L}^T \Phi_f \textbf{L})\\
& \text{subject to}
& & \X = \textbf{L} + \M, \\
\end{aligned}
\label{eq:rpca-ug-step1}
\end{equation}

In the second step, $\textbf{L}$ and $\M$ are fixed and we solve the following optimization problem with respect to $\Phi_f$:
\begin{equation}
\begin{aligned}
& \underset{\Phi_f}{\text{minimize}}
& & \gamma tr(\textbf{L}^T \Phi_f \textbf{L}) + \beta \|\Phi_f \|_F^2 \\
& \text{subject to}
& & \Phi_f \in \mathcal{L} \\
\end{aligned}
\label{eq:rpca-ug-step2}
\end{equation}
We also use ADMM for (\ref{eq:rpca-ug-step2}).  We reformulate (\ref{eq:rpca-ug-step2}) as:
\begin{equation}
\begin{aligned}
& \underset{\Phi_f}{\text{minimize}}
& & \gamma tr(\textbf{L}^T\Phi_f\textbf{L})+\beta\|\Phi_f\|_F^2\\
& \text{subject to}
& & \Phi_f -\z = 0, \\
&&& \z \in \mathcal{L}
\end{aligned}
\label{eq:rpca-ug-step2-ADMM}
\end{equation}
where $\z \in \R^{p \times p}$ is the auxiliary variables matrix.
The augmented Lagrangian of (\ref{eq:rpca-ug-step2-ADMM}) is:
\begin{equation}
\begin{aligned}
L_\rho(\Phi_f,\z,\u) = &\gamma tr(\textbf{L}^T\Phi_f\textbf{L})+\beta\|\Phi_f\|_F^2 \\
& +\frac{\rho}{2}\|\z-\Phi_f\|_F^2 + \langle \u,\z-\Phi_f \rangle \\
\end{aligned}
\label{eq:rpca-ug-step2-Lagrangian}
\end{equation}
where $\u \in \R^{p \times p}$ is the dual variables matrix. And we use the following formulas to update  $\Phi_f$, $\z$ and $\u$:

\begin{equation}
\begin{aligned}
& \Phi_f^{(k+1)} := \frac{\rho \z^{(k)} - \gamma \textbf{L}^T\textbf{L} + \u^{(k)}}{\frac{\beta}{2} +\rho}\\
& \z^{(k+1)}:= \prod_{\mathcal{L}}(\Phi_f^{(k+1)} - \frac{1}{\rho}\u^{(k)})\\
& \u^{(k+1)}:= \u^{(k)} + \frac{1}{k}(\z^{(k+1)}-\Phi_f^{(k+1)}) \\
\end{aligned}
\label{eq:rpca-ug-step2-update}
\end{equation}
where $\rho>0$ is the Lagrangian parameter and $\prod_{\mathcal{L}}$ is the Euclidean projection onto set $\mathcal{L}$.
In what follows, we provide an analysis and experiment results of this alternating minimization algorithm.


\section{Impact of inexact graph}
\label{sec:SensitivityAnalysis}

In the first step of our proposed algorithm, i.e. (\ref{eq:rpca-ug-step1}), we estimate the low-rank components $\textbf{L}$ by leveraging the graph Laplacian $\Phi_f$. In many problems, the estimated graph can be noisy and distorted. Intuitively, the effectiveness of step 1 (low-rank components estimation) depends on the quality of the estimated graph. In particular, our hypothesis is that: the better is the quality of the estimated graph, the more effective will be this graph-assisted low-rank components estimation.  This hypothesis also motivates our alternating optimization framework, which refines the graph in the second step in order to achieve better low-rank components estimation in the first step.

In what follows, we perform an analysis to quantify the effectiveness of graph-assisted low-rank components estimation (the first step) under different distorted graph conditions.  As we will discuss, this analysis validates our hypothesis in a rigorous manner. It sheds some light on how the quality of the estimated graph impacts the effectiveness of low-rank components estimation. The analysis also provides insight on settings of some parameters in the formulation, in particular, $\gamma$ in (\ref{eq:rpca-ug-step1}), which is an important parameter to determine the weight of graph regularization.

To quantify the effectiveness of the first step, we use the following measurement.
Given $\textbf{L}^{(k)}$, the low-rank components estimation at the $k$-th iteration within the first step (recall that the first step is solved iteratively using ADMM), we derive an analytic expression for this quantity (under different graph distortion):
\begin{equation}
\begin{aligned}
L_{err} \triangleq \|{\textbf{L}^{(k+2)}} - \textbf{L}_0\|_F
\label{eq:LowrankErrorFunc}
\end{aligned}
\end{equation}
Here $\textbf{L}_0$ is the ground-truth low-rank components.
In other words, starting from a certain $\textbf{L}^{(k)}$, we quantify the improvement in the low-rank components estimation  after two iterations  
under different distortion of the graph.  
One would expect that if the quality of the graph is good, low-rank components estimation error will be much reduced after two iterations, i.e., $L_{err}$ will be small. 
Note that the mathematical formulas become rather clumsy beyond two iterations.  Therefore, we choose to quantify the  improvement  after two iterations under various graph distortion.


To derive an analytic expression to characterize $L_{err}$, we proceed as follows:
\begin{enumerate}
\item Given ${\textbf{L}^{(k)}}$  we derive the formula for ${\textbf{L}^{(k+2)}}$; 
\item We derive an approximated expression for $\|\textbf{L}^{(k+2)}\|_F$;
\item We derive an approximated expression  for $L_{err}$ based on 
the approximated $\|\textbf{L}^{(k+2)}\|_F$
 and analyze it under different graph distortion.
 \end{enumerate}

\subsection{Mathematical expression for ${\textbf{L}^{(k+2)}}$}
\label{ssec:deductL_k+2}

In the first step, we apply ADMM to solve (\ref{eq:rpca-ug-step1}). Based on the ADMM, the equation (\ref{eq:rpca-ug-step1}) is re-formulated as:
\begin{equation}
\begin{aligned}
& \underset{\textbf{L},\M}{\text{minimize}}
& & \|\textbf{L}\|_* + \delta \|\M\|_1 +   \gamma tr(\K^T \Phi_f \K)\\
& \text{subject to}
& & \X = \textbf{L} + \M, \\
& & & \textbf{L} = \K \\
\end{aligned}
\label{eq:rpca-ug-step1_transform}
\end{equation}

(\ref{eq:rpca-ug-step1_transform}) is solved iteratively as follows:
\begin{align*}
\J^{(k+1)} & := \frac{r_1(\X-\M^{(k)}+\frac{\z_1^{(k)}}{r_1})+r_2(\K^{(k)}+\frac{\z_2^{(k)}}{r_2})}{r_1+r_2}\\
\textbf{L}^{(k+1)} & := D_{\tau_1}(\J^{(k+1)})\\
\M^{(k+1)} & := \Omega_{\tau_2}(\X-\textbf{L}^{(k+1)}+\frac{\z_1^{(k)}}{r_1})\\
\K^{(k+1)} & := r_2(\gamma\Phi_f + r_2\I)^{-1}(\textbf{L}^{(k+1)} - \frac{\z_2^{(k)}}{r_2}) \stepcounter{equation}\tag{\theequation}\label{eq:ADMM_deduction}\\
\z^{(k+1)}_1 & := \z_1^{(k)} + r_1(\X-\textbf{L}^{(k+1)}-\M^{(k+1)})\\
\z^{(k+1)}_2 & := \z_2^{(k)} + r_2(\K^{(k+1)}-\textbf{L}^{(k+1)})\\
\end{align*}
Here $D_{\tau_1}(\J^{(k+1)}) \triangleq \U\Omega_{\tau_1}(\Sigma)\V^T$ and $\J^{(k+1)} = \U\Sigma\V^T$ is the singular value decomposition (SVD) of $\J^{(k+1)}$. $\Omega_{\tau_2}(x) \triangleq sgn(x)max(|x|-\tau_2,0)$ is a element-wise soft-thresholding operator. $\z_1$ and $\z_2$ are dual variables with updating step size $r_1$ and $r_2$ separately. And $\tau_1 = \frac{2}{r_1+r_2}$ and $\tau_2 = \frac{\delta}{r_1}$. 

We assume that the graph is inexact: $\tilde{\Phi_f}= \Phi_{f0} + \Delta\Phi_f$. Given $\textbf{L}^{(k)},\M^{(k)},\K^{(k)},\z_1^{(k)},\z_2^{(k)}$, the $\J^{(k+1)},\textbf{L}^{(k+1)},\M^{(k+1)}$ and $\z_1^{(k+1)}$ 
have not been  influenced by the graph distortion 
as shown in the steps in  (\ref{eq:ADMM_deduction}). Meanwhile, 
\begin{equation}
\K^{(k+1)} = r_2\lbrack\gamma(\Phi_{f0} + \Delta\Phi_f) + r_2\I\rbrack^{-1}(\textbf{L}^{(k+1)} - \frac{\z_2^{(k)}}{r_2})
\label{eq:K_k+1}
\end{equation} 
is affected by the distorted graph. 

For the $(k+2)$-th iteration, $\J^{(k+2)}$ is also influenced by the graph distortion: 
\begin{align*}
\J^{(k+2)} = & \frac{r_1(\X-\M^{(k+1)}+\frac{\z_1^{(k+1)}}{r_1})}{r_1+r_2} + \frac{r_2(\K^{(k+1)}+\frac{\z_2^{(k+1)}}{r_2})}{r_1+r_2}\\
= & \frac{r_1\X-r_1\M^{(k+1)}+ \z_1^{(k)} + r_1\X-r_1\textbf{L}^{(k+1)}-r_1\M^{(k+1)}}{r_1+r_2}\\
& + \frac{r_2\K^{(k+1)}+\z_2^{(k)} + r_2\K^{(k+1)}-r_2\textbf{L}^{(k+1)}}{r_1+r_2}\\
= & \frac{2r_1\X - 2r_1\M^{(k+1)} + 2r_2\K^{(k+1)}}{r_1+r_2}\\
& - \frac{(r_1+r_2)\textbf{L}^{(k+1)}+\z_1^{(k)} +\z_2^{(k)}}{r_1+r_2} \stepcounter{equation}\tag{\theequation}\label{eq:J_k+2}\\
= & \frac{2r_1}{r_1+r_2}(\X-\M^{(k+1)}) + \frac{2r_2}{r_1+r_2}\K^{(k+1)}\\
& - \textbf{L}^{(k+1)} -\frac{\z_1^{(k)} +\z_2^{(k)}}{r_1+r_2} \\
= & \frac{2r_1}{r_1+r_2}(\X-\M^{(k+1)}) -\textbf{L}^{(k+1)} -\frac{\z_1^{(k)} +\z_2^{(k)}}{r_1+r_2}\\
& +\frac{2r_2^2}{r_1+r_2}\cdot\lbrack\gamma(\Phi_{f0} + \Delta\Phi_f) + r_2\I\rbrack^{-1}(\textbf{L}^{(k+1)} - \frac{\z_2^{(k)}}{r_2})\\
= & \A + \lbrack\gamma(\Phi_{f0} + \Delta\Phi_f) + r_2\I\rbrack^{-1}\cdot \B\\
\end{align*}
where $ \A = \frac{2r_1}{r_1+r_2}(\X-\M^{(k+1)}) -\textbf{L}^{(k+1)} -\frac{\z_1^{(k)} +\z_2^{(k)}}{r_1+r_2} \in \R^{p \times n}$ and $\B = \frac{2r_2^2}{r_1 + r_2} \cdot (\textbf{L}^{(k+1)} -\frac{\z_2^{(k)}}{r_2}) \in \R^{p \times n}$.

Therefore, $\textbf{L}^{(k+2)}$ can be obtained as:
\begin{equation}
\begin{aligned}
\textbf{L}^{(k+2)} =  & D_{\tau_1}(\J^{(k+2)})\\
= & D_{\tau_1} \lbrace\A + \lbrack\gamma(\Phi_{f0} + \Delta\Phi_f) + r_2\I\rbrack^{-1}\cdot \B\rbrace
\end{aligned}
\label{eq:L_k+2}
\end{equation}



\subsection{An approximated expression for $\|\textbf{L}^{(k+2)}\|_F$}
\label{ssec:approx_Lerr}
The expression in (\ref{eq:L_k+2}) for computing $\textbf{L}^{(k+2)}$ requires SVD in the $D_{\tau_1}(.)$ operator and is rather difficult to analyze. Here we present a simplified expression for  $\|\textbf{L}^{(k+2)}\|_F$, which is used later for approximating $L_{err}$.
We recall that 
$\textbf{L}^{(k+2)} = D_{\tau_1}(\J^{(k+2)})$ where $\tau_1 = \frac{2}{r_1+r_2}$.
\begin{align*}
& \|D_{\tau_1}(\J^{(k+2)})\|^2_F \\
= &\sum_{i=1}^p\Omega_{\tau_1}^2(\sigma_i)\\
= &\sum_{i=1}^p\sigma_i^2 - \tau_1\sum_{i=1}^p (2\sigma_i-\tau_1)\cdot \mathbbm{1}(\sigma_i>\tau_1)\\
& -\sum_{i=1}^p \sigma_i^2\cdot \mathbbm{1}(\sigma_i \leq \tau_1)\\
= &\|\J^{(k+2)}\|_F^2 - 2\tau_1\sum_{i=1}^p \sigma_i\cdot \mathbbm{1}(\sigma_i >\tau_1) \\
& - 2\tau_1\sum_{i=1}^p \sigma_i\cdot \mathbbm{1}(\sigma_i \leq \tau_1) + \sum_{i=1}^p\tau_1^2\cdot \mathbbm{1}(\sigma_i >\tau_1) \\
& + 2\tau_1\sum_{i=1}^p \sigma_i\cdot \mathbbm{1}(\sigma_i \leq\tau_1) -\sum_{i=1}^p \sigma_i^2\cdot \mathbbm{1}(\sigma_i \leq \tau_1)\\
= &\|\J^{(k+2)}\|_F^2 - 2\tau_1\|\J^{(k+2)}\|_*  + \tau_1^2\cdot rank(\textbf{L}^{(k+2)}) \\
& + \sum_{i=1}^p(2\tau_1-\sigma_i)\sigma_i\cdot \mathbbm{1}(\sigma_i\leq \tau_1)\stepcounter{equation}\tag{\theequation}\label{eq:L_k+2_F_approx}\\
\overset{\text{(1)}}{\approx} & \|\J^{(k+2)}\|_F^2 - 2\tau_1\|\J^{(k+2)}\|_*  \\
& + \tau_1^2\cdot rank(\textbf{L}^{(k+2)}) + \tau_1\sum_{i=1}^p\sigma_i\cdot \mathbbm{1}(\sigma_i\leq \tau_1)\\
\overset{\text{(2)}}{\approx} & \|\J^{(k+2)}\|_F^2 - 2\tau_1\|\J^{(k+2)}\|_* \\
& +\tau_1^2\cdot rank(\textbf{L}^{(k)}) + \tau_1\cdot \frac{\tau_1}{h}\cdot \lbrack p-rank(\textbf{L}^{(k)})\rbrack
\end{align*}
where $p$ is the number of dimensionality of $\Sigma$ in $\J^{(k+1)} = \U\Sigma\V^T$ and $h= \frac{\tau_1}{ave(\sum_{i=1}^p\sigma_i\cdot \mathbbm{1}(\sigma_i\leq \tau_1))}$. For the approximation $\overset{\text{(1)}}{\approx}$ in the function (\ref{eq:L_k+2_F_approx}), because $\sigma_i \leq \tau_1$ is true for the term $\sum_{i=1}^p(2\tau_1-\sigma_i)\sigma_i \cdot\mathbbm{1}(\sigma_i\leq\tau_1)$, one has $2\tau_1-\sigma_i = \tau_1 +(\tau_1 -\sigma_i) \geq \tau_1 \approx \tau_1$ and $\sum_{i=1}^p(2\tau_1-\sigma_i)\sigma_i\cdot \mathbbm{1}(\sigma_i\leq \tau_1) \approx \tau_1\sum_{i=1}^p\sigma_i\cdot \mathbbm{1}(\sigma_i\leq \tau_1)$. There are two approximations applied in the $\overset{\text{(2)}}{\approx}$ step of the function (\ref{eq:L_k+2_F_approx}). The first one is $rank(\textbf{L}^{(k+2)}) \approx rank(\textbf{L}^{(k)})$. Because the matrix rank between neighboring $\textbf{L}^{(k)}$ has small difference. The second approximation is $\sum_{i=1}^p\sigma_i\cdot \mathbbm{1}(\sigma_i\leq \tau_1) \approx \frac{\tau_1}{h}\cdot \lbrack p-rank(\textbf{L}^{(k)})\rbrack$ where $ \frac{\tau_1}{h} \approx ave(\sum_{i=1}^p\sigma_i\cdot \mathbbm{1}(\sigma_i\leq \tau_1))$. 
Thus, the approximation of $\|\textbf{L}^{(k+2)}\|^2_F$ is:
\begin{equation}
\begin{aligned}
\|{\textbf{L}^{(k+2)}}\|^2_F = & \|D_{\tau_1}({\J^{(k+2)}})\|^2_F\\
\approx & \|{\J^{(k+2)}}\|_F^2 - 2\tau_1\|{\J^{(k+2)}}\|_* \\
& +\tau_1^2\cdot rank(\textbf{L}^{(k)}) + \tau_1\cdot \frac{\tau_1}{h}\cdot \lbrack p-rank(\textbf{L}^{(k)})\rbrack
\end{aligned}
\label{eq:L_k+2_approx}
\end{equation}
where ${\J^{(k+2)}} = \A + \lbrack\gamma(\Phi_{f0} + \Delta\Phi_f) + r_2\I\rbrack^{-1}\cdot \B$. 



\begin{figure}[htbp]\small
\centering
\includegraphics[width=0.7\columnwidth]{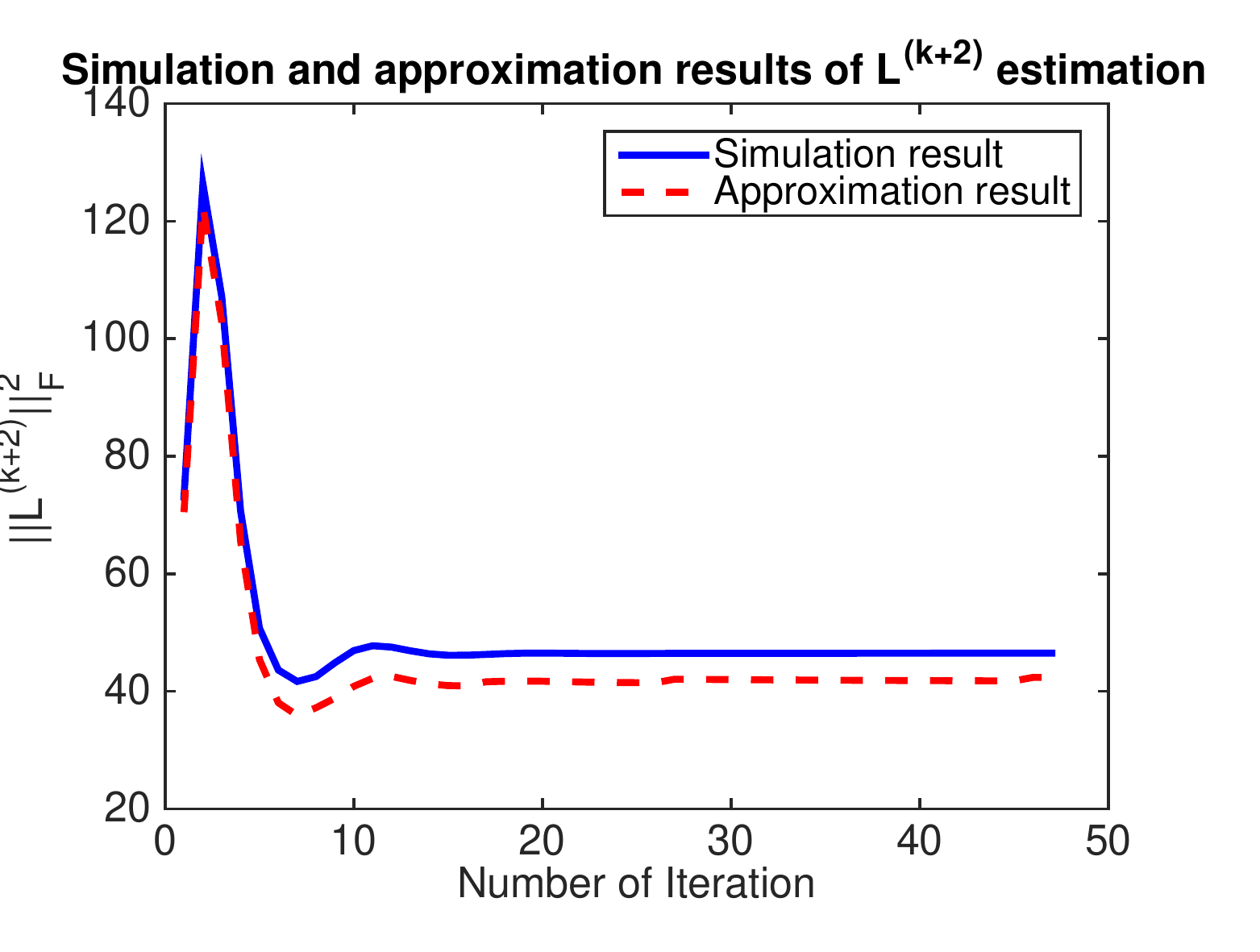} \tabularnewline
\caption{Comparison between the empirical values obtained during simulation and the approximation of $\|\textbf{L}^{(k+2)}\|^2_F$ using (\ref{eq:L_k+2_approx}), for certain $L^{(k)}$. We depict the comparison for different $k$ (number of iteration).}
\label{fig:Simulation_VS_Analytical_L_k+2}
\end{figure}

Figure \ref{fig:Simulation_VS_Analytical_L_k+2} depicts the comparison of some empirical values of $\|\textbf{L}^{(k+2)}\|^2_F$ obtained during the simulation and the  approximations calculated by function (\ref{eq:L_k+2_approx}). As shown in  the figure, the approximation results are quiet close to the empirical ones.
Thus, we use expression (\ref{eq:L_k+2_approx}) to approximate $\|\textbf{L}^{(k+2)}\|^2_F$ later in calculating $L_{err}$, and this avoids SVD of $\J^{(k+2)}$. 

We further simplify 
$\|\textbf{L}^{(k+2)}\|^2_F$.
We aim to illustrate how the graph distortion $\Delta \Phi_f$
affects $\|\textbf{L}^{(k+2)}\|^2_F$ and subsequently $L_{err}$.
We first simplify ${\J^{(k+2)}}$.
Here we use two approximations: \\
1) If $\epsilon $ is a small number, then $(\J +\epsilon\X)^{-1} = \J^{-1} -\epsilon\J^{-1}\X\J^{-1} + O(\epsilon^2)$\cite{henderson1981deriving}. Then, we can obtain: 
\begin{equation}
\begin{aligned}
& \lbrack\gamma(\Phi_{f0} + \Delta\Phi_f) +r_2\I\rbrack ^{-1}\\
= &(\gamma\Phi_{f0}+r_2\I)^{-1} -\epsilon(\gamma\Phi_{f0}+r_2\I)^{-1}(\gamma\hat{\Phi_{f0}})(\gamma\Phi_{f0}+r_2\I)^{-1} \\
& + O(\epsilon^2)\\
\approx & (\gamma\Phi_{f0}+r_2\I)^{-1}-\epsilon(\gamma\Phi_{f0}+r_2\I)^{-1}(\gamma\hat{\Phi_{f0}})(\gamma\Phi_{f0}+r_2\I)^{-1}\\
\end{aligned}
\label{eq:MatrixSummationSeparation_Approximation}
\end{equation} 
2) We apply $(\I+\J)^{-1} = \I -\J+\J^{2} -\J^{3}+\cdots$. Then, $(\gamma\Phi_{f0}+r_2\I)^{-1}$ can be approximated as:
\begin{equation}
\begin{aligned}
(\gamma\Phi_{f0}+r_2\I)^{-1} = & \lbrack r_2(\frac{\gamma}{r_2}\Phi_{f0} + \I)\rbrack^{-1}\\
 \approx & \frac{1}{r_2}\lbrack \I -\frac{\gamma}{r_2}\Phi_{f0} + (\frac{\gamma}{r_2}\Phi_{f0})^{2}\rbrack \\
\end{aligned}
\label{eq:MatrixInverse_Approximation}
\end{equation}
Therefore, we have:
\begin{equation}
\begin{aligned}
{\J^{(k+2)}} = & \A + \lbrack \gamma(\Phi_{f0}+\Delta \Phi_f) + r_2\I\rbrack^{-1}\cdot \B \\
\approx  & \A + \lbrack\C - \C(\gamma \Delta \Phi_f)\C\rbrack\cdot \B \\
= & \A + \C\B - \gamma\C\Delta\Phi_f\C\B \\
\end{aligned}
\label{eq:Distorted_J_k+2_estimation_approx}
\end{equation}
where $\C = \frac{1}{r_2}\lbrack \I -\frac{\gamma}{r_2}\Phi_{f_0} + (\frac{\gamma}{r_2}\Phi_{f_0})^2\rbrack \in \R^{p \times p}$. 
Based on  (\ref{eq:L_k+2_approx}) and (\ref{eq:Distorted_J_k+2_estimation_approx}), we obtain:
\begin{equation}
\begin{aligned}
\|\textbf{L}^{(k+2)}\|^2_F \approx & \|\A + \C\B - \gamma\C\Delta\Phi_f\C\B\|_F^2 \\
& - 2\tau_1\|\A + \C\B - \gamma\C\Delta\Phi_f\C\B\|_* \\
& + \tau_1^2\cdot rank(\textbf{L}^{(k)}) + \tau_1\cdot \frac{\tau_1}{h}\cdot \lbrack p -rank(\textbf{L}^{(k)})\rbrack\\
\end{aligned}
\label{eq:Distorted_L_approx1}
\end{equation}

To further simplify $\|\cdot\|_*$, we apply an inequality\cite{srebro2004learning} between $\|\cdot\|^2_F$ and $\|\cdot\|_*$ :
\begin{equation}
\begin{aligned}
\|{\J^{(k+2)}}\|_*\leq &\sqrt{rank({\J^{(k+2)}})\|{\J^{(k+2)}}\|^2_F}\\
\approx & g\sqrt{rank({\J^{(k+2)}})}\|{\J^{(k+2)}}\|_F \\
\label{eq:nuclear_approx}
\end{aligned}
\end{equation}
where $g$ is calculated  from empirical data. Thus, combine (\ref{eq:Distorted_L_approx1}) and (\ref{eq:nuclear_approx}), $\|\textbf{L}^{(k+2)}\|_F$  can be approximated as: 
\begin{equation}
\begin{aligned}
\|\textbf{L}^{(k+2)}\|_F \approx & \lbrace\|\A + \C\B - \gamma\C\Delta\Phi_f\C\B\|_F^2 \\
& - 2\tau_1g\sqrt{p}\cdot\|\A + \C\B - \gamma\C\Delta\Phi_f\C\B\|_F \\
& + \tau_1^2\cdot rank(\textbf{L}^{(k)}) + \tau_1\cdot \frac{\tau_1}{h}\cdot \lbrack p-rank(\textbf{L}^{(k)})\rbrack \rbrace^{\frac{1}{2}}\\
\label{eq:Distorted_L_approx3}
\end{aligned}
\end{equation}

\subsection{Analytical expression for $L_{err}$}
As discussed in \ref{eq:LowrankErrorFunc} and according to the reverse triangle inequality, we have: 
\begin{equation}
\begin{aligned}
L_{err} \triangleq \|{\textbf{L}^{(k+2)}} - \textbf{L}_0\|_F\approx |\|{\textbf{L}^{(k+2)}}\|_F - \|\textbf{L}_0\|_F|
\label{eq:LowrankErrorFunc101}
\end{aligned}
\end{equation}
Combine (\ref{eq:Distorted_L_approx3}) and (\ref{eq:LowrankErrorFunc101}), we have:
\begin{equation}
\begin{aligned}
L_{err} \approx & |\|{\textbf{L}^{(k+2)}}\|_F - \|\textbf{L}_0\|_F|\\
\approx &|(\|\A + \C\B - \gamma\C\Delta \Phi_f\C\B\|_F^2\\
& - 2\tau_1g\sqrt{p}\cdot\|\A + \C\B -\gamma\C\Delta \Phi_f\C\B\|_F + m)^{\frac{1}{2}}\\
&- \|\textbf{L}_0\|_F|\\
\end{aligned}
\label{eq:LowrankErrorFunc_approx}
\end{equation}
Here, 
$\A$, $\B$ and $\C$ are independent of $\Delta \Phi_f$. And $m = \tau_1^2\cdot rank(\textbf{L}^{(k)}) + \tau_1\cdot \frac{\tau_1}{h}\cdot \lbrack p-rank(\textbf{L}^{(k)})\rbrack$ is a constant independent of $\Delta \Phi_f$.

\begin{figure*}[htbp]
\footnotesize
\begin{centering}
\begin{tabular}[t]{c c c c}
Original setting & Change rank($\textbf{L}_0$)& Change \#nodes in $\mathcal{G}$ & Change \#samples \tabularnewline
$r=6,p=30,n=50$ & $r= 3,p=30,n=50$ & $r=6,p=10,n=50$ &$r=6,p=30,n=30$\tabularnewline
\includegraphics[width=0.49\columnwidth]{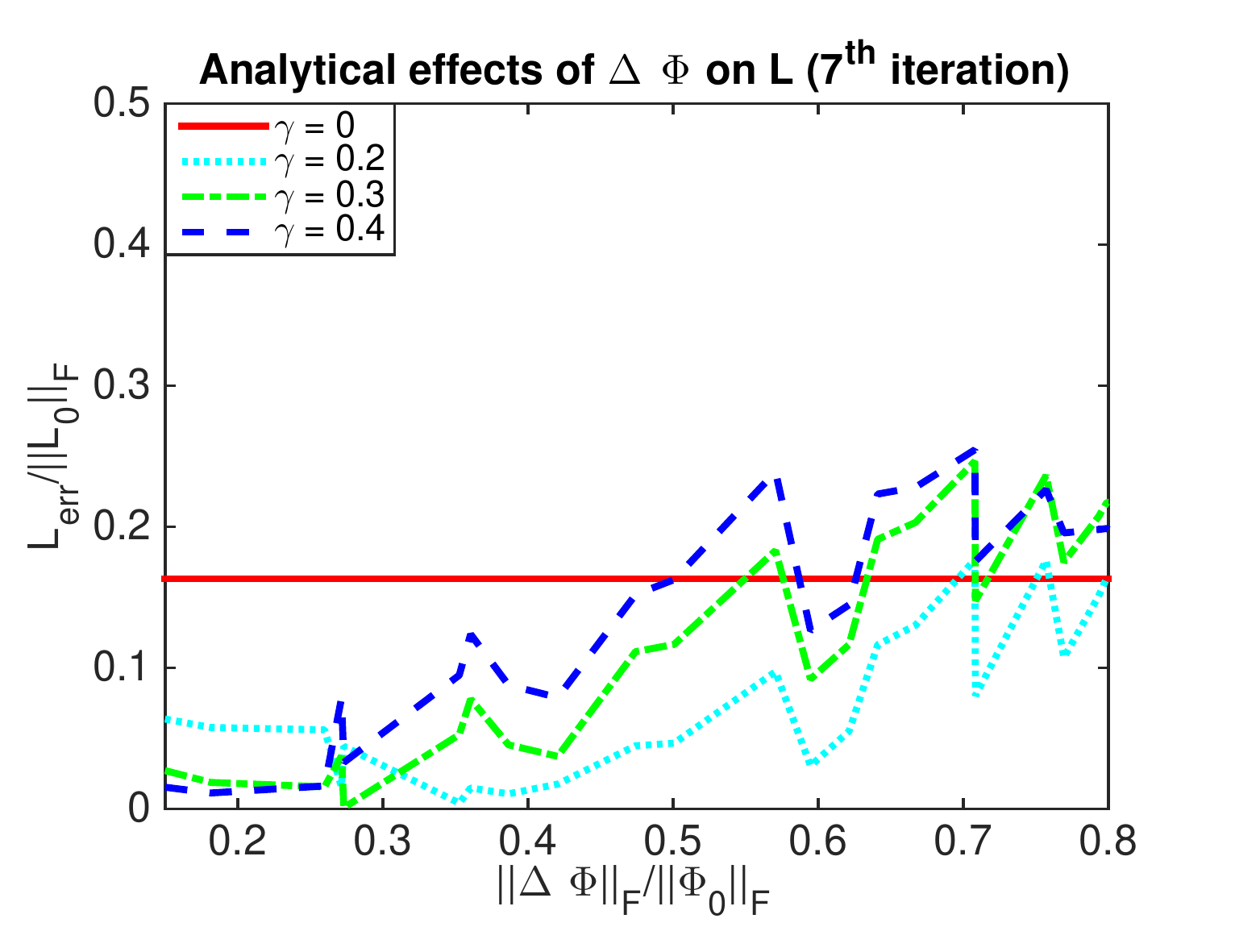} &
\includegraphics[width=0.49\columnwidth]{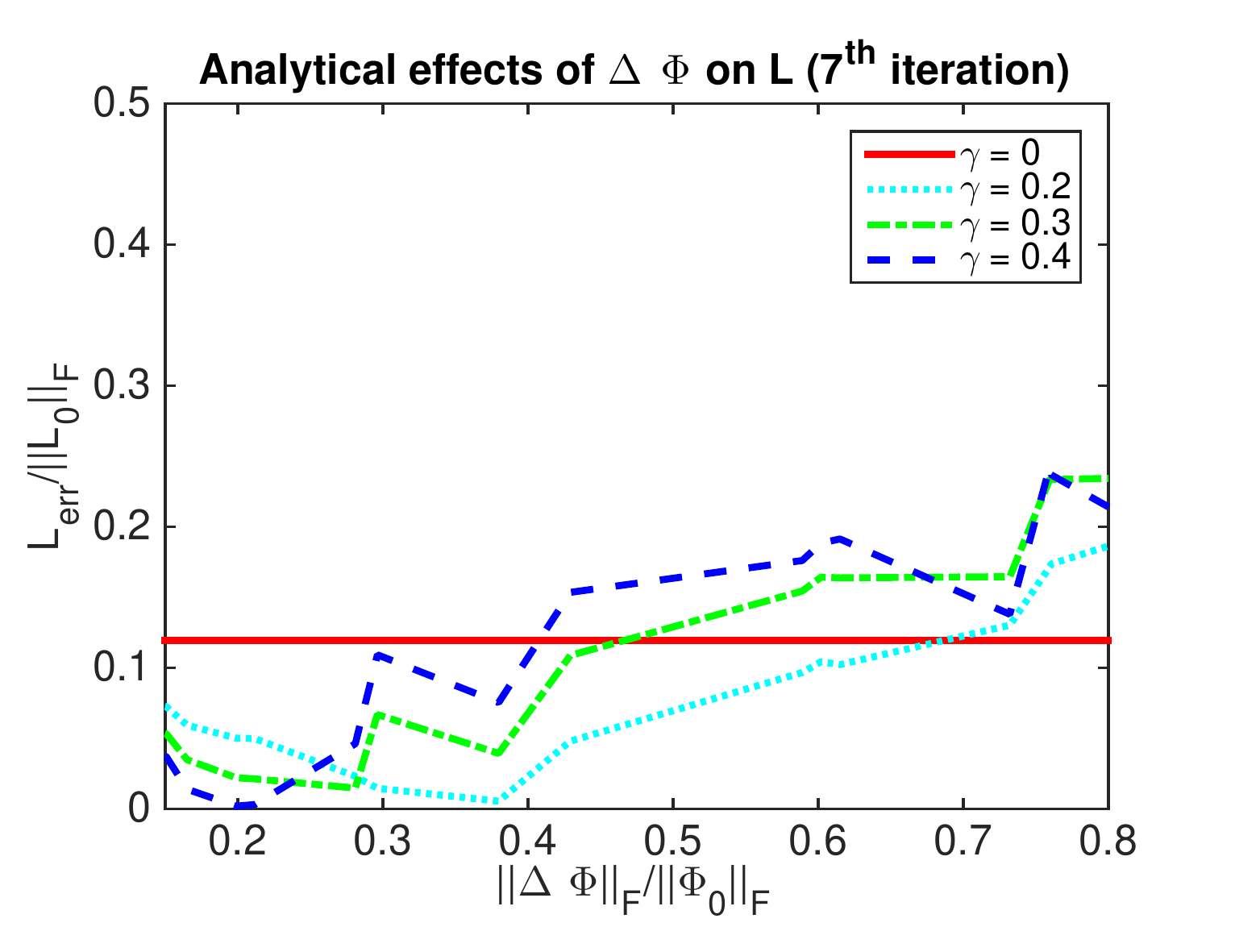} &
\includegraphics[width=0.49\columnwidth]{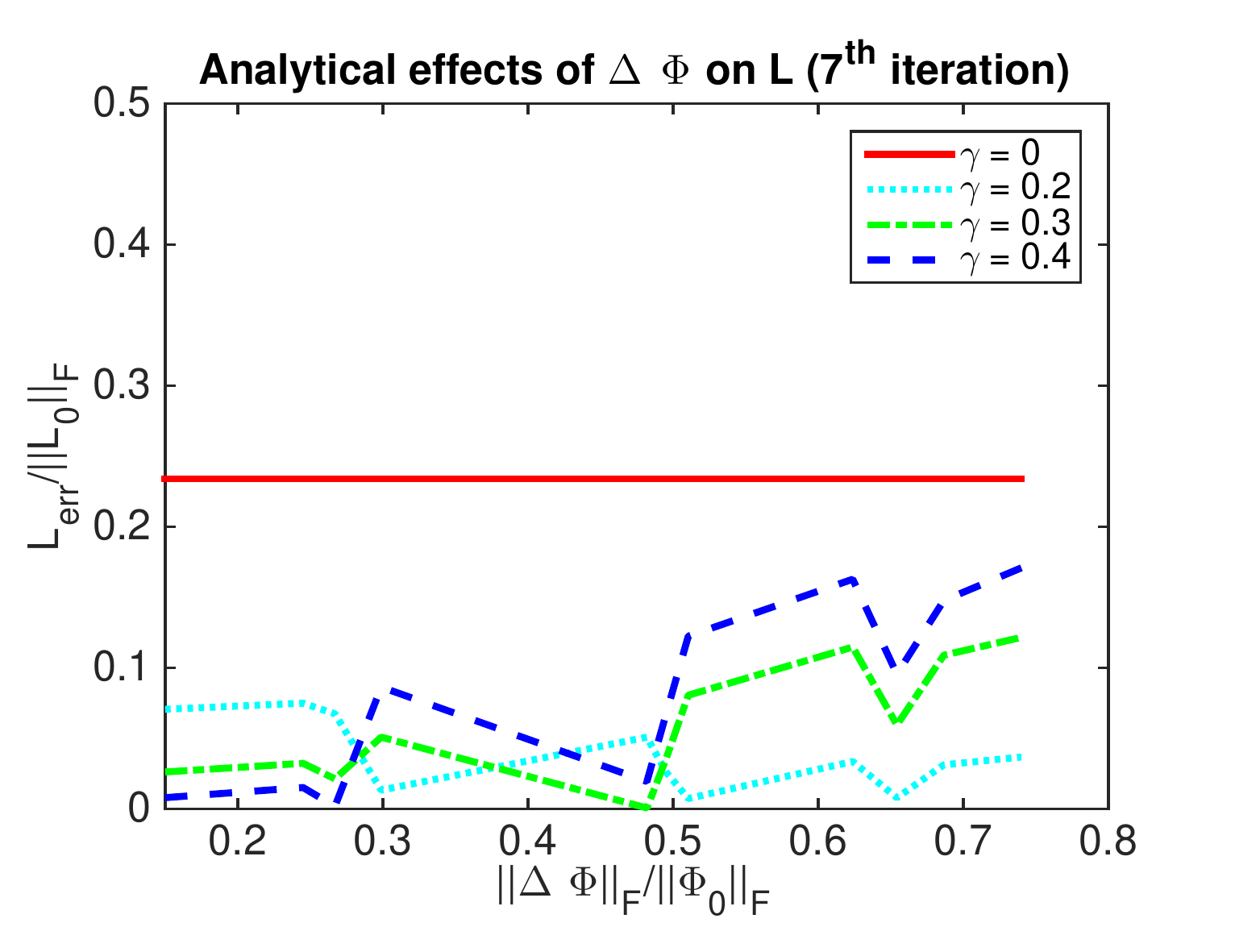} & 
\includegraphics[width=0.49\columnwidth]{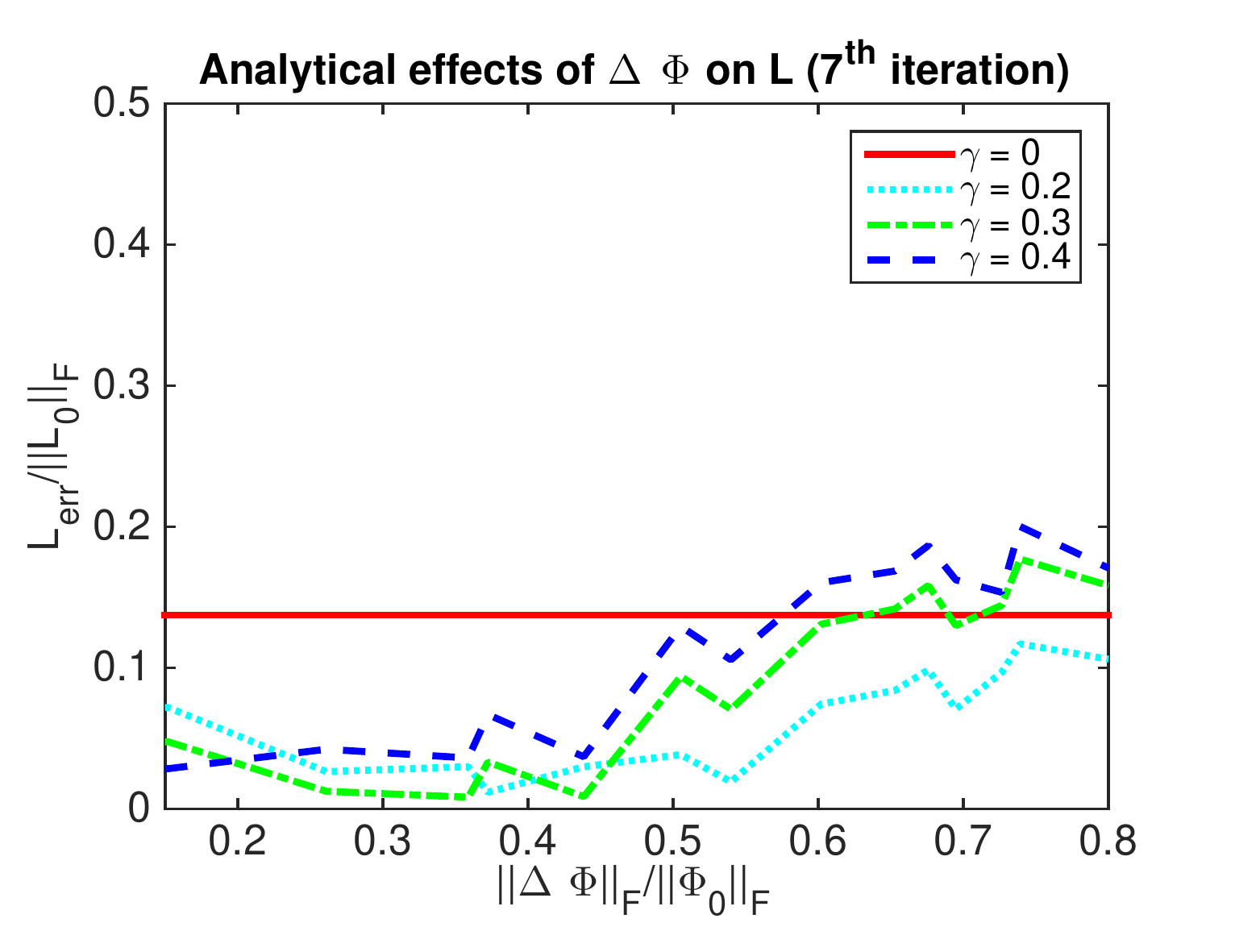} \tabularnewline  
\includegraphics[width=0.49\columnwidth]{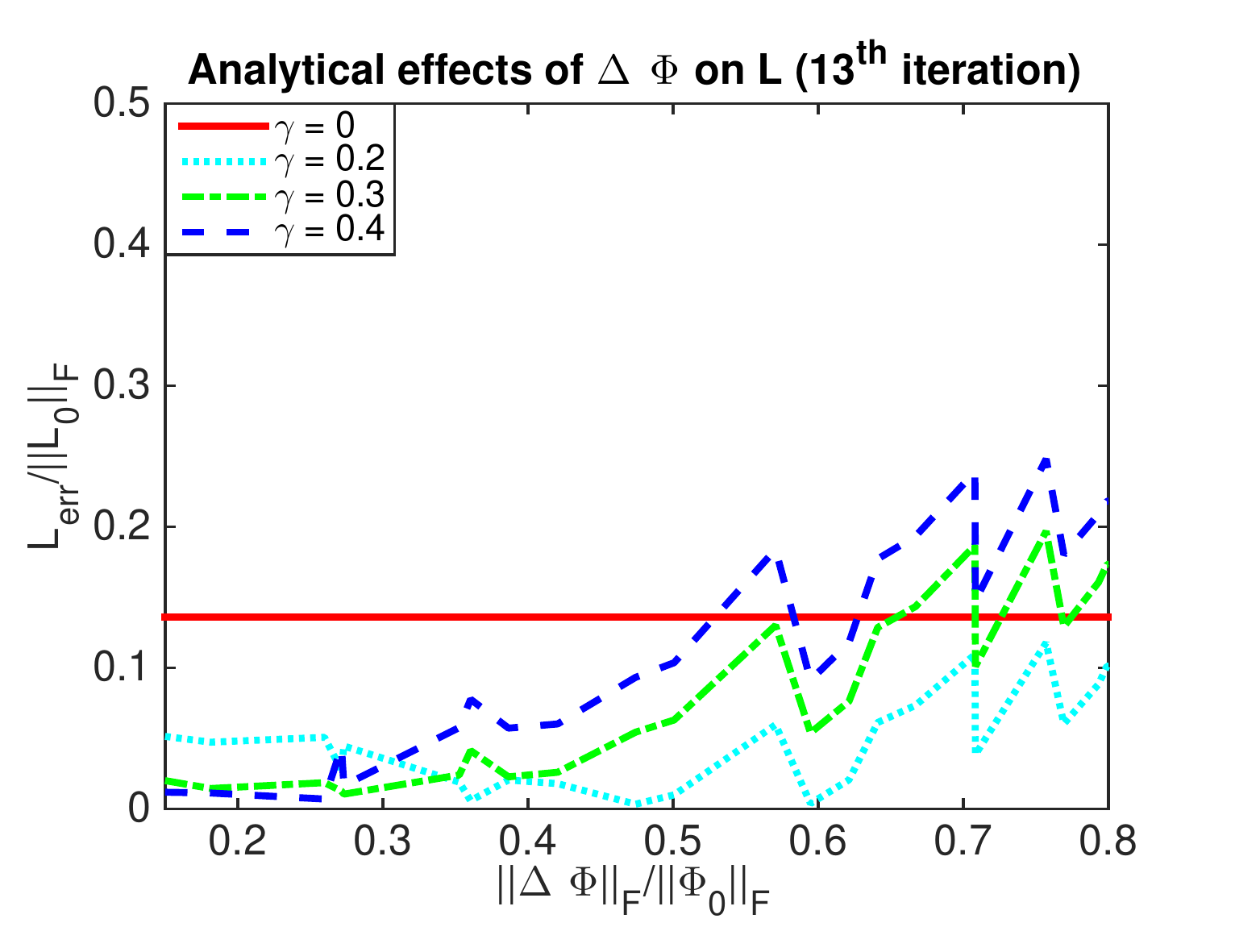} &
\includegraphics[width=0.49\columnwidth]{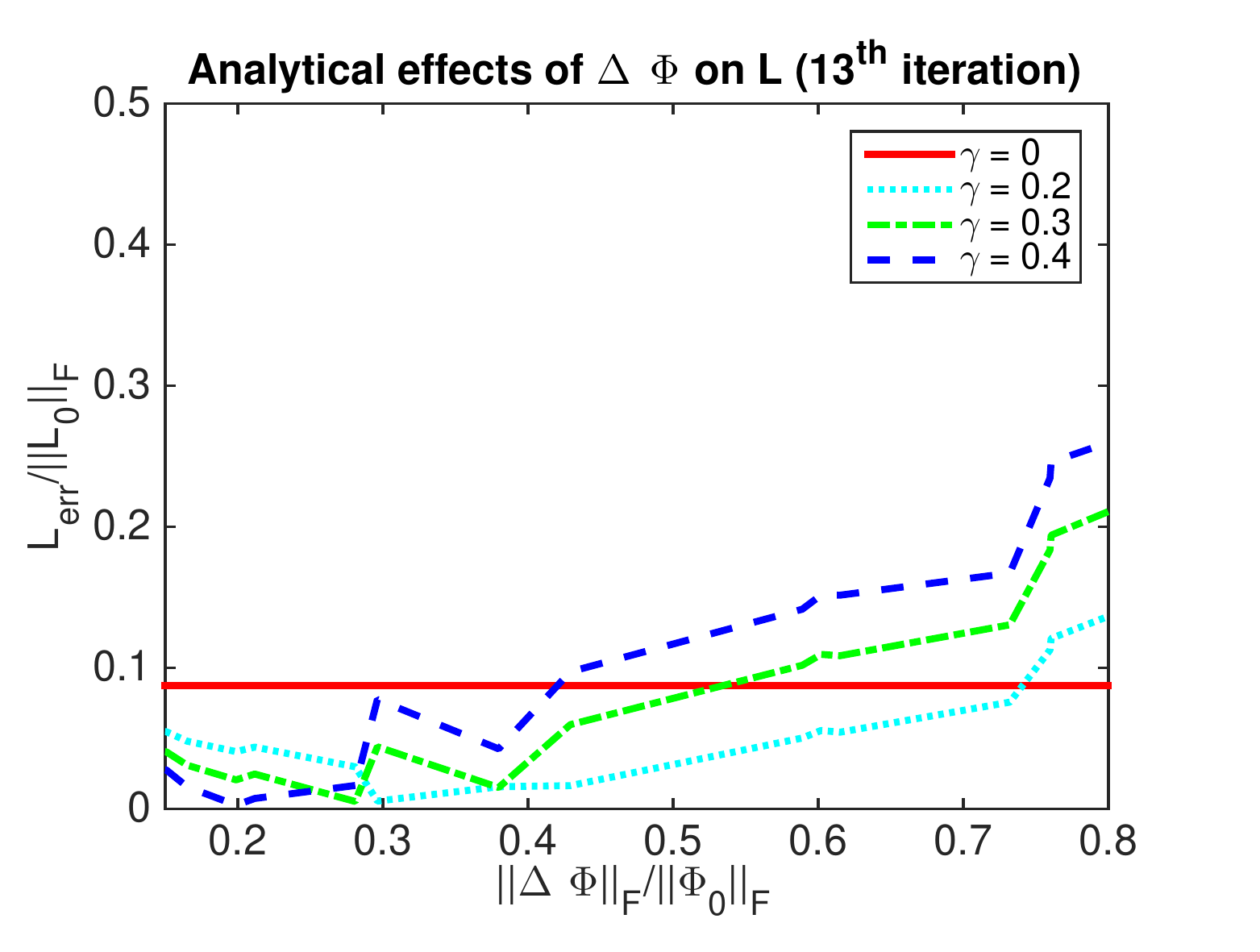} &
\includegraphics[width=0.49\columnwidth]{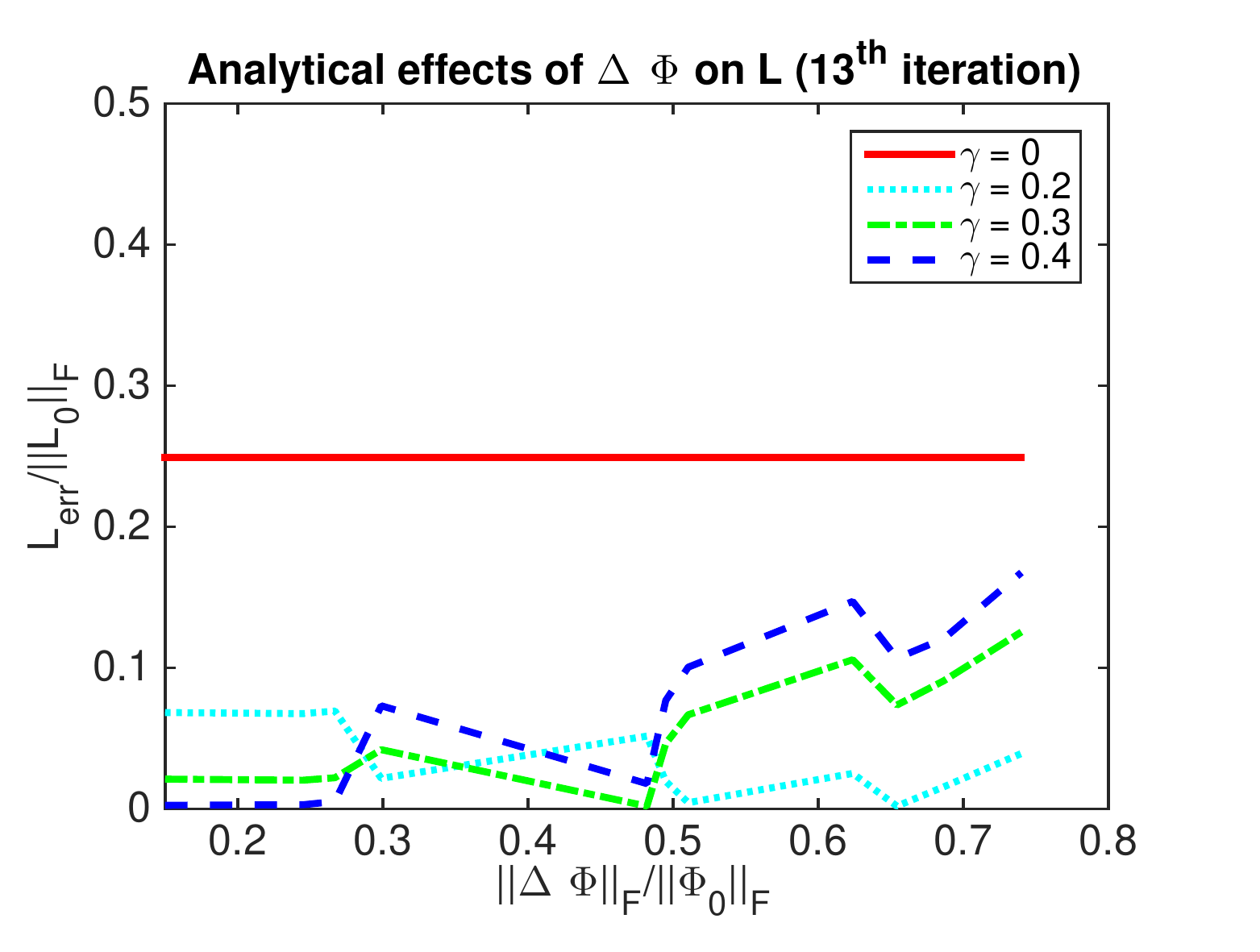} & 
\includegraphics[width=0.49\columnwidth]{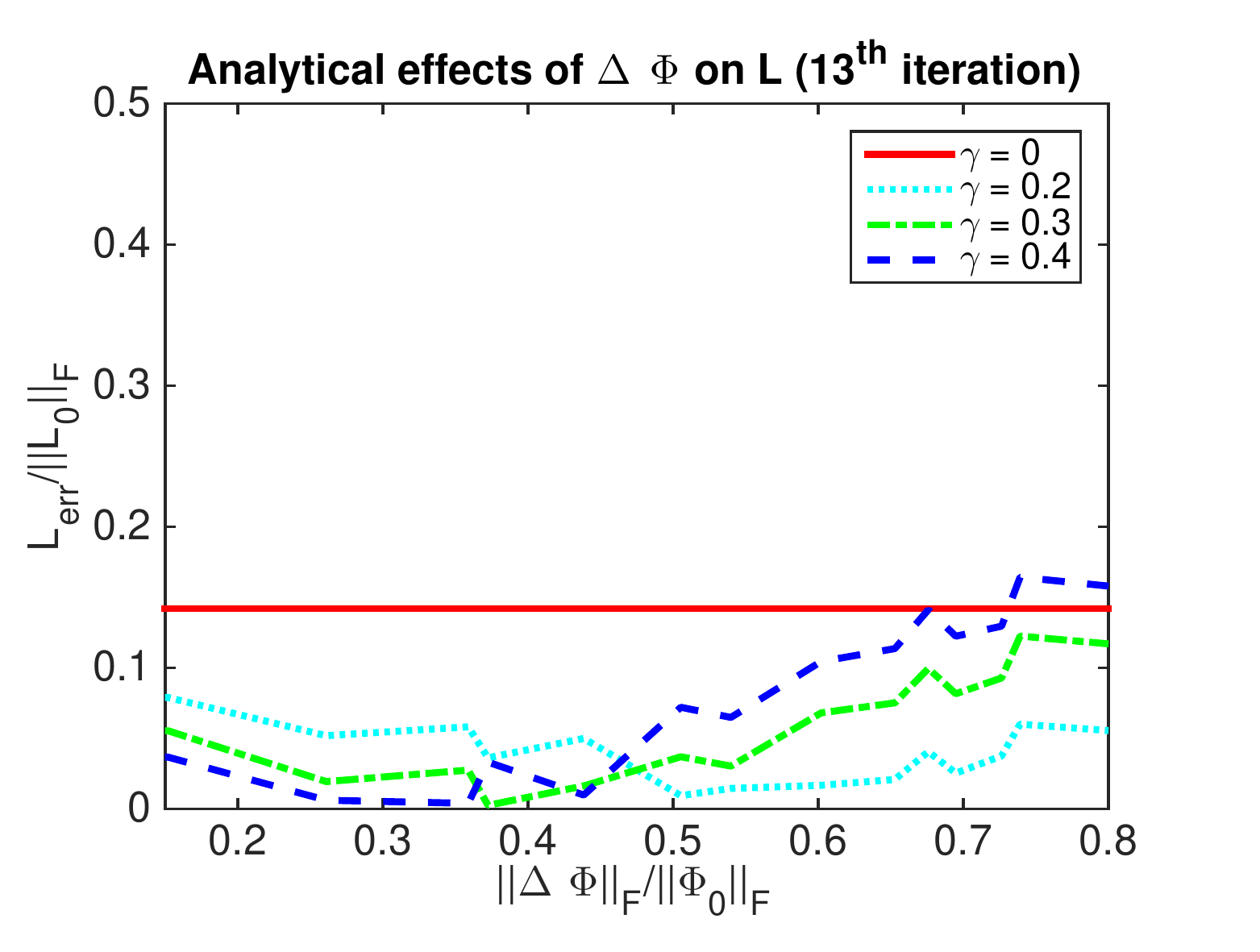} \tabularnewline
\includegraphics[width=0.49\columnwidth]{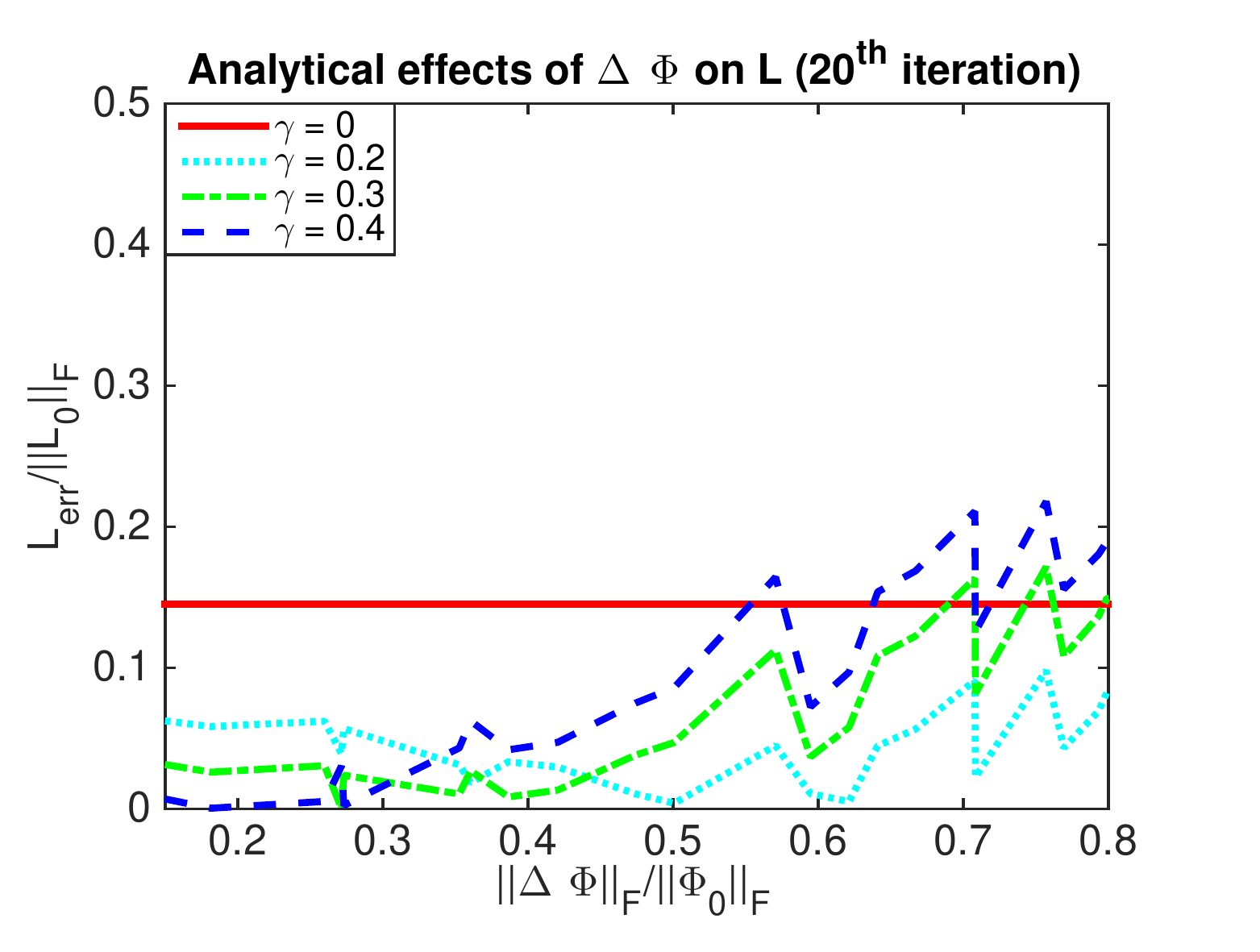} &
\includegraphics[width=0.49\columnwidth]{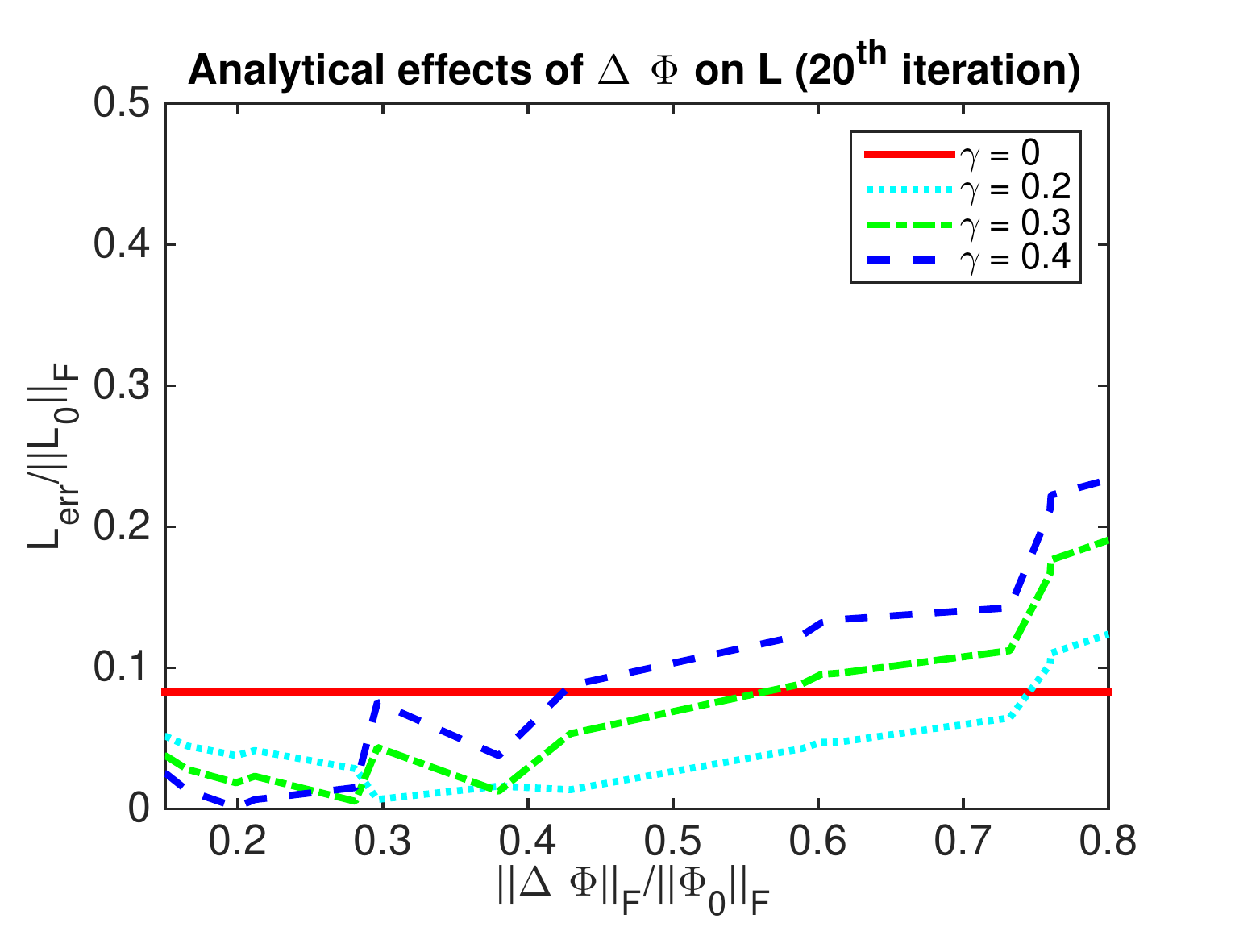} &
\includegraphics[width=0.49\columnwidth]{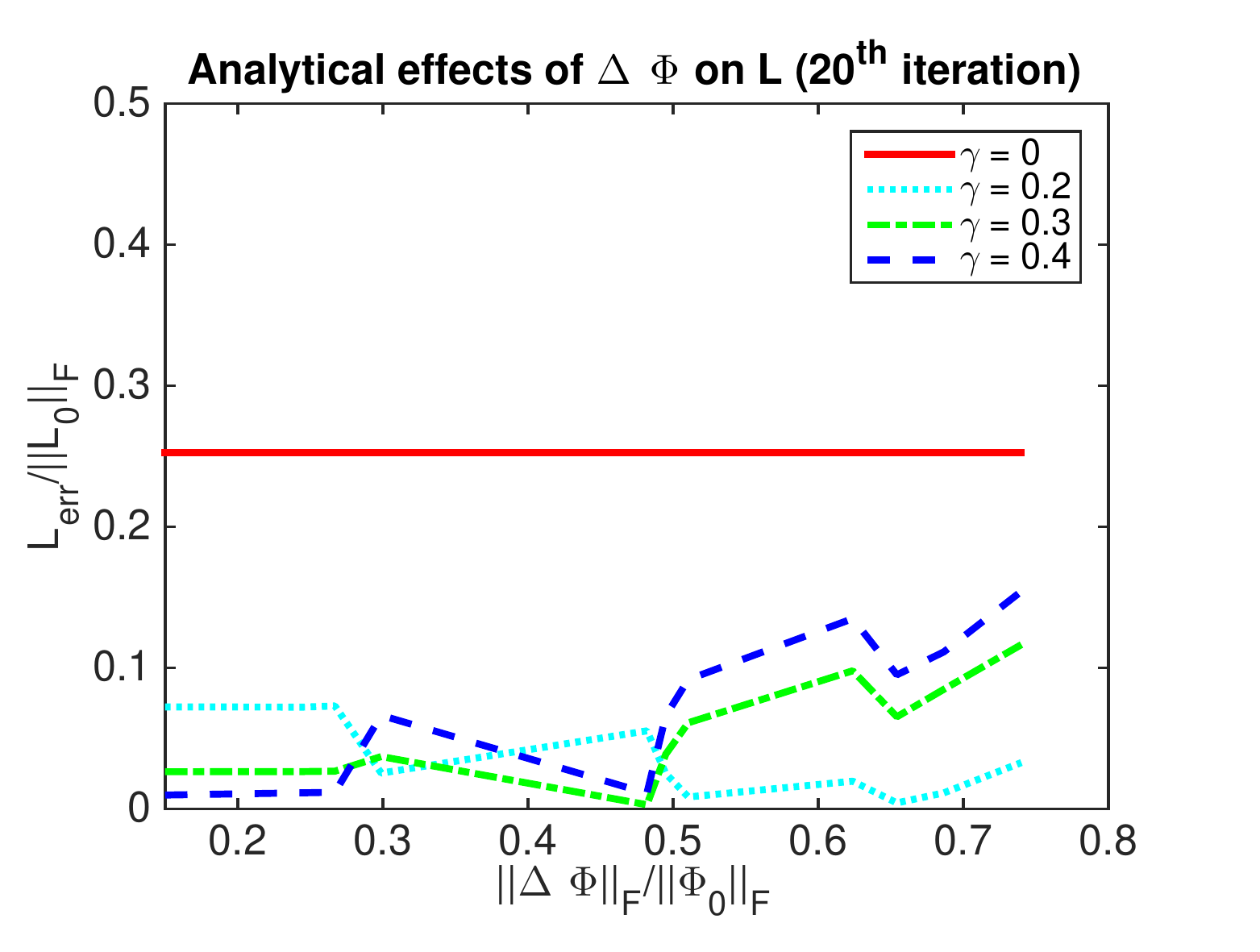} & 
\includegraphics[width=0.49\columnwidth]{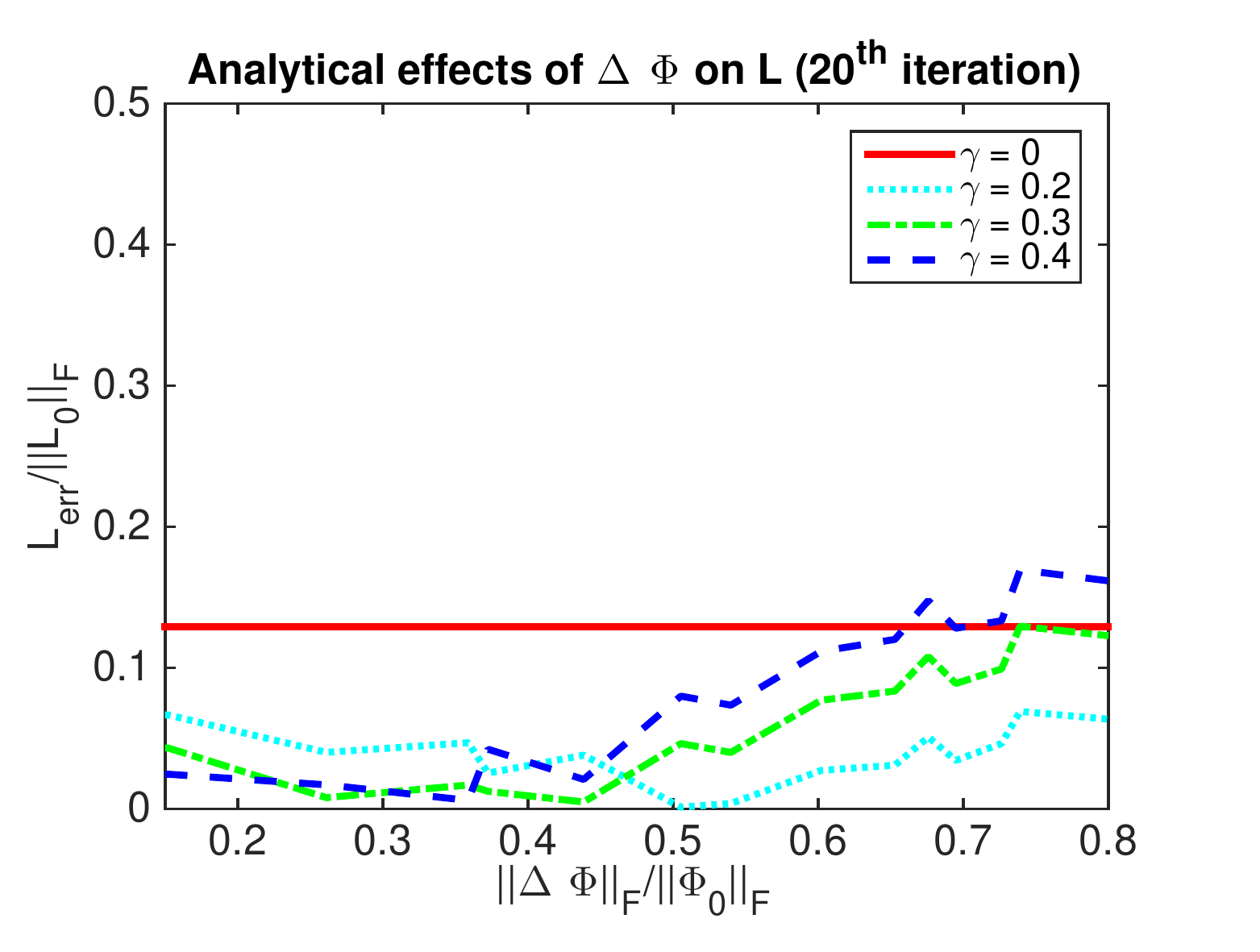} \tabularnewline
\end{tabular}
\par\end{centering}
\caption{Analysis of the effect of graph distortion $\Delta \Phi_f$ on the estimation of low-rank components $\textbf{L}$ under different settings. For the results in the first column, the input data is generated with 6 components extracted from the graph Laplacian matrix of a 30-node graph (i.e., the intrinsic dimensionality is 6) with 50 samples. For the results in the second column, the input data is generated with 3 components extracted from a 30-node graph Laplacian matrix with  50 samples. For the results in the third column, the input data is generated with 6 components extracted from a 10-node graph Laplacian matrix with  50 samples. And for the results in the last column, the input data is generated with 6 components extracted from a 30-node graph Laplacian matrix with 30 samples. Different rows depict  $\frac{L_{err}}{\|\textbf{L}_0\|_F}$ 
at several different iterations ($7^{th}$, $13^{th}$ and $20^{th}$ iterations).}
\label{fig:Nov15_Simulation_VS_Analytical}
\end{figure*}

\subsection{Effect of distorted graphs}
\label{ssec:SA_result}

Based on the approximation in (\ref{eq:LowrankErrorFunc_approx}),
we analyze the effect of the distorted graphs on 
$L_{err}$.  
In this analysis, the data matrix $\X$, ground-truth low-rank components $\textbf{L}_0$ and ground-truth graph $\Phi_{f0}$ are generated using the procedure in Section \ref{sec:Synthetic}. 
Then we apply the first step of LGE (See function (\ref{eq:rpca-ug-step1})) with $\Phi_{f0}$ and $\delta = 0.5$.
We vary the amount of distortion in the graphs as follows:
the distortion follows an i.i.d. Bernoulli distribution with a probability $s$ and the amplitude of distortion follows a uniform distribution $U(0,1)$. We change $s$ from 0.01 to 1 with 0.01 as step size. 
We perform analysis with graph distortion from $\frac{\|\Delta \Phi_f\|_F}{\|\Phi_0\|_F} = 15\%$ to $\frac{\|\Delta \Phi_f\|_F}{\|\Phi_0\|_F} = 80\%$.
We also vary $\gamma$, the weight of the graph regularization (See function (\ref{eq:rpca-ug-step1})), to understand the interaction between the weight of the graph regularization and the accuracy of low-rank components estimation under different graph distortion.
We perform analysis on different settings:
different rank $r$ of $\textbf{L}_0$, number of nodes $p$ in the graph $\mathcal{G}$ as well as number of samples $n$. 
Figure \ref{fig:Nov15_Simulation_VS_Analytical} depicts the analysis results at  3 different iterations of the first step of LGE: $7^{th}$, $13^{th}$ and $20^{th}$ iterations (i.e., $k$ = 5, 11, 18 respectively).


In Figure~\ref{fig:Nov15_Simulation_VS_Analytical}, the solid (red) line is the plot with $\gamma=0$, which means that the graph is not involved in the estimation of the low-rank components. 
Other dotted lines correspond to different values of $\gamma$. They indicate different weights in the graph regularization in function (\ref{eq:rpca-ug-step1}) and therefore different extents to which the graph is leveraged in the low-rank components estimation.  From the analytical results in Figure~\ref{fig:Nov15_Simulation_VS_Analytical}, we observe that:
\begin{itemize}
\item The use of graph regularization leads to smaller low-rank components estimation errors provided that the graph distortion is not large.  This can be observed in Figure~\ref{fig:Nov15_Simulation_VS_Analytical}: the plots with $\gamma > 0$ have smaller $\frac{L_{err}}{\|\textbf{L}_0\|_F}$ compared to the plot with $\gamma = 0$ when graph distortion $\|\Delta \Phi_f\|_F$ is not large.
\item For all $\gamma$, there is a general trend that as the graph distortion increases, the low-rank components estimation error increases and hence the graph-assisted low-rank components estimation (in the first step) becomes less effective.  In particular, when the graph distortion is large, the use of graph has detrimental effect on the low-rank components estimation, as we can observe that when $\|\Delta \Phi_f\|_F$ is large,
the plots with 
$\gamma > 0$ have larger $\frac{L_{err}}{\|\textbf{L}_0\|_F}$ compared to the plot with $\gamma = 0$ in figure~\ref{fig:Nov15_Simulation_VS_Analytical}.
\item When $\|\Delta \Phi_f\|_F$ is small, we can achieve better low-rank components estimation with a large $\gamma$. On the other hand, when $\|\Delta \Phi_f\|_F$ is large, the use of large $\gamma$ has  negative impact on the low-rank components estimation. 
\end{itemize}
Overall, our  analysis suggests that in general there are advantages in leveraging the graph in the low-rank components estimation, even when there exists distortion in the graph. However, there are subtle issues that one needs to consider, as we have discussed. 
Our proposed LGE alternates between low-rank components estimation and graph refinement to improve the effectiveness of graph-assisted low-rank components estimation.  In the next section, we provide experiment results based on both synthetic and real data to further demonstrate the effectiveness.



\section{Experiment with synthetic data}
\label{sec:Synthetic}
In the previous section, we analyzed the performance of the first step of our proposed LGE (i.e., graph assisted low-rank estimation) under different graph distortion. 
In this and next sections, we provide experimental results of LGE using synthetic and real data.
For synthetic data, 
we generate some low-rank, graph-smooth and grossly-corrupted dataset by the following model:
\begin{equation}
\begin{aligned}
\X = \textbf{L}_0 + \M
\end{aligned}
\label{eq:data-model}
\end{equation}
where $\textbf{L}_0\in \R^{p\times n}$ is a ground-truth low-rank components matrix with rank $r$ and $\M$ is the sparse matrix. 

The low-rank matrix is a product $\textbf{L}_0 = \P\Y^T$ where $\P\in \R^{p\times r}$ contains basis vectors and $\Y\in \R^{n\times r}$ is the coefficient matrix. 
$\textbf{L}_0$ is smooth on a graph  built with the following steps.
The (ground-truth) graph $\mathcal{G}$ consists of $p$ nodes, with each pair of nodes having a probability of $q$ to be connected together. The edge weights between connected nodes are drew uniformly from 0 to 1 and presented in a $p\times p$ symmetric adjacency matrix $\W$. 
Then we can calculate the Laplacian matrix $\Phi_f$ from $\W$ and compute the eigenvectors and eigenvalues of $\Phi_f$. As small eigenvalues correspond to eigenvectors smooth on the graph, we select the eigenvectors corresponding to smallest $r$ eigenvalues as the basis vectors, i.e., the columns of $\P$. For coefficient matrix $\Y$, the entries are independently sampled from a $N(\mu,1/p)$ distribution. 
For the perturbation matrix $\M$, we introduce the error following an i.i.d Bernoulli distribution with a probability $d= \|M\|_0 / pn$. 
The amplitude of distorted elements are computed with different methods as discussed later.

We perform several experiments on the synthetic data to illustrate some properties of LGE.
In these experiments, synthetic data are generated with the following settings: for low-rank components matrix $\textbf{L}_0$, a $p=30$ nodes graph $\mathcal{G}$ with $\mu=0, q =0.3$ is built, its graph  Laplacian matrix $\Phi_{f0}$ is calculated, and $r=3$ eigenvectors of $\Phi_{f0}$ are chosen to be the columns of $\P$.  Therefore, $\textbf{L}_0$ is graph-smooth with  rank 3. And $n=50$ samples are generated to obtain $\textbf{L}_0\in\R^{30\times 50}$. For sparse matrix $\M$, we generate different matrices suitable for different experiments.


\subsection{Experiments on individual steps of LGE}
\label{ssec:AnalysisEachStep}
LGE updates low-rank components estimation and graph estimation iteratively. As discussed in Section~\ref{sec:lowrank_GraphEstimate}, these two steps enhance  each other to obtain the final optimal result.
In what follows, we report experiment results to illustrate how the two steps interact.

\subsubsection{Step 1: better graph improves low-rank components estimation}
\label{sssec:AnalysisStep1}

In this experiment, we focus on the first step to understand the impact of the input graph $\Phi_f$ on the  output low-rank components $\textbf{L}$.  With distortion on the graph: $\tilde{\Phi_f} = \Phi_{f0} + \Delta \Phi_f$, the low-rank components estimation, $\hat{\textbf{L}}$, is affected. The estimation error is measured as: $\| \hat{\textbf{L}} -\textbf{L}_0\|_F/\|\textbf{L}_0\|_F$. 

\begin{figure}\small
\begin{centering}
\begin{tabular}{cc}
\includegraphics[width=0.49\columnwidth]{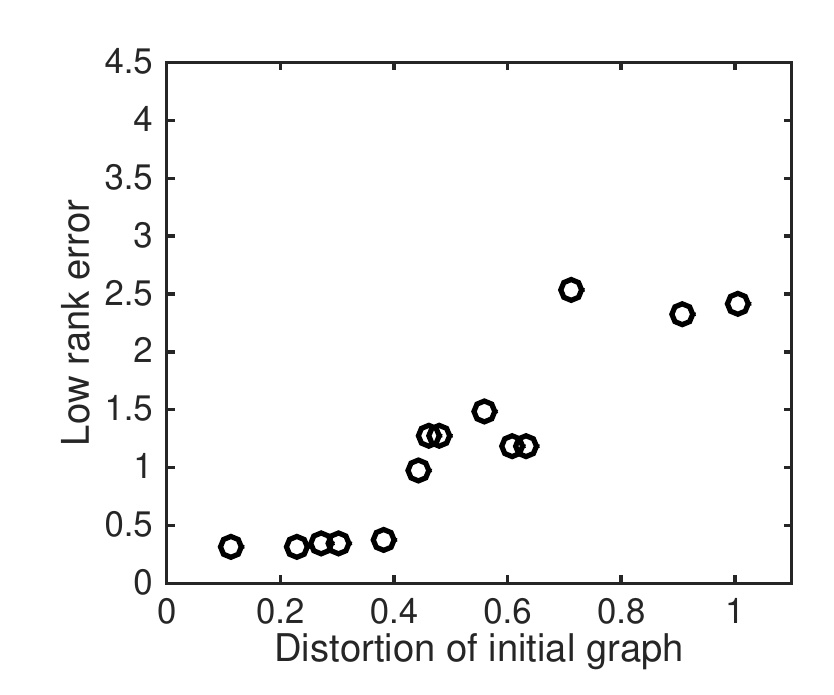} & 
\includegraphics[width=0.49\columnwidth]{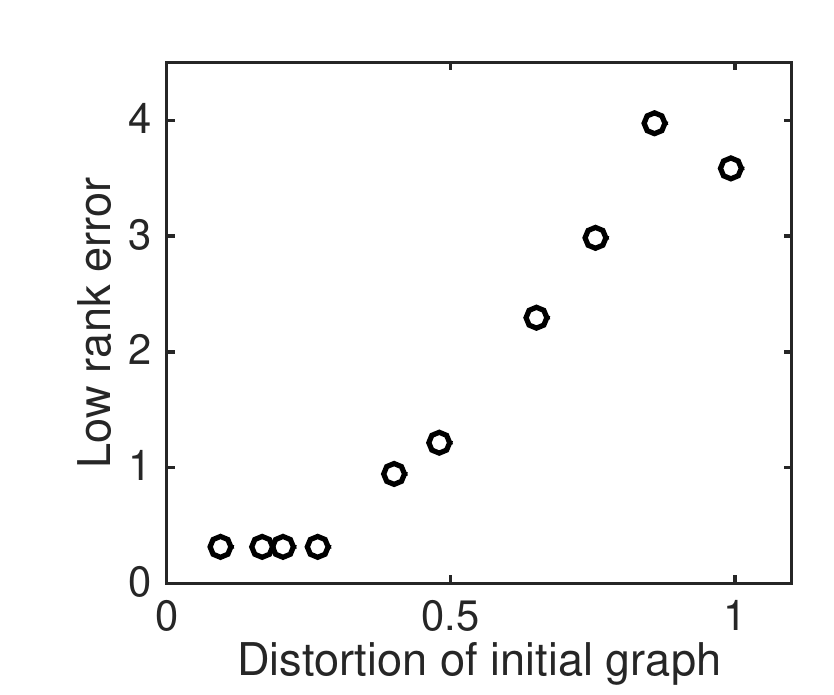}\tabularnewline
(a) & (b) \tabularnewline
\end{tabular}
\par\end{centering}
\caption{Low-rank components estimation error under different distortions on the  initial graph. (a) The topology of initial graph remains unchanged. Only the edge weights are distorted. (b) The topology of initial graph changes with the distortion, and edges can be added or deleted.}
\label{fig:Step1Result}
\end{figure}

We experiment with two kinds of distortion $\Delta \Phi_f$. First, the topology of graph remains the same. In particular, only non-zero off-diagonal elements (non-zero elements correspond to existing edges) of $\Phi_{f0}$ are added with distortion which follows a Bernoulli distribution with probability $s$. The amplitudes of distorted elements follow a uniform distribution $U(0,1)$.  Second, the graph topology changes as the result of the added distortion. In particular, all off-diagonal elements are added with distortion following a Bernoulli distribution with a probability of $s$. Therefore,  new edges may be introduced in the graph. The amplitudes of elements in $\Delta \Phi_f$ follow a uniform distribution $U(0,1)$. Thus, $s$ controls the distortion percentage on the initial graph $\tilde{\Phi_f}$.  For topology-unchanged distortion, we vary $s$ from 0.02 to 0.65. For topology-changed distortion, we vary $s$  from 0.05 to 0.3. 
We measure the low-rank components estimation error $\| \hat{\textbf{L}} -\textbf{L}_0\|_F/\|\textbf{L}_0\|_F$ under these distortions on the graph.
Figure \ref{fig:Step1Result} depicts how the distortion in the input graph affects the output.
The horizontal axis depicts graph distortion 
$\| \tilde{\Phi_f} -\Phi_{f0}\|_F/\|\Phi_{f0}\|_F$ resulted from certain $s$.
We use the mean value of five runs as the result for each setting.
The results suggest that a more precise graph leads to better low-rank components estimation. Furthermore, the estimation error is more severe when distortion causes   changes in graph topology. 

\subsubsection{Step 2: better low-rank components results in better graph estimation}
\label{sssec:AnalysisStep2}

In this experiment, we focus on the second step of LGE to understand how the low-rank components affect the graph estimation. By adding distortion of different  sparsity to the original low-rank components $\tilde{\textbf{L}}=\textbf{L}_0 + \Delta \textbf{L}$, the graph estimation, $\hat{\Phi_f}$,  is affected. We measure the error in graph estimation as: $\| \hat{\Phi_f} -\Phi_{f0}\|_F/\|\Phi_{f0}\|_F$. 

We experiment two ways to generate distortion matrix $\Delta \textbf{L}$. First, distorted elements follow a Bernoulli distribution with a probability $u= \|\Delta \textbf{L}\|_0 / pn$.  Therefore, $u$ controls the distortion sparsity. The amplitudes of the distorted elements are $\pm1$ with equal probability. 
Note that the amplitudes of the distortion are rather significant (in comparison, the average amplitude of $\textbf{L}_0$ equals to 0.15). Second, distorted elements follow a Bernoulli distribution with a probability $u= \|\Delta \textbf{L}\|_0 / pn$, but the amplitudes of distorted elements follow a uniform distribution $U(0,c)$. 

\begin{figure}\small
\begin{centering}
\begin{tabular}{cc}
\includegraphics[width=0.49\columnwidth]{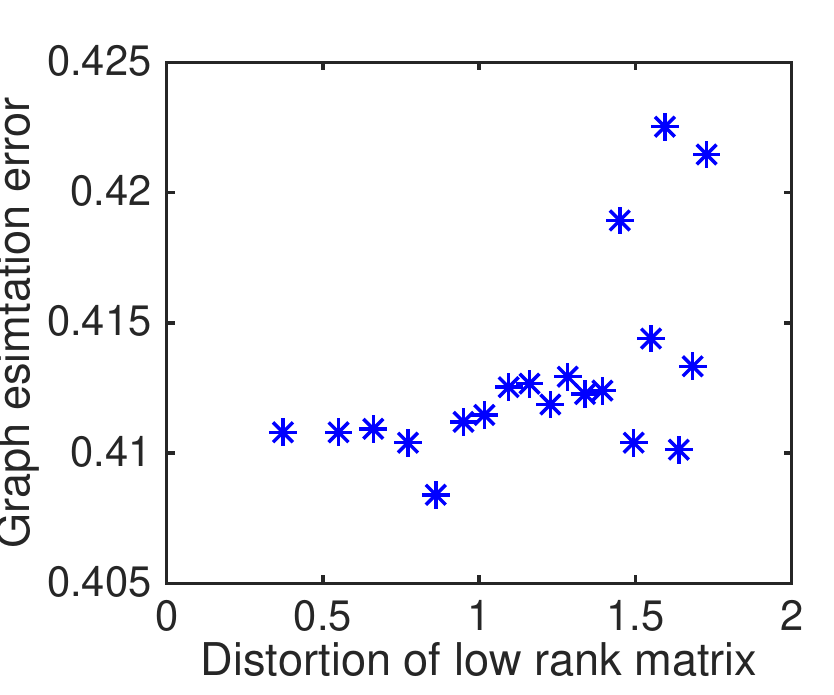} & 
\includegraphics[width=0.49\columnwidth]{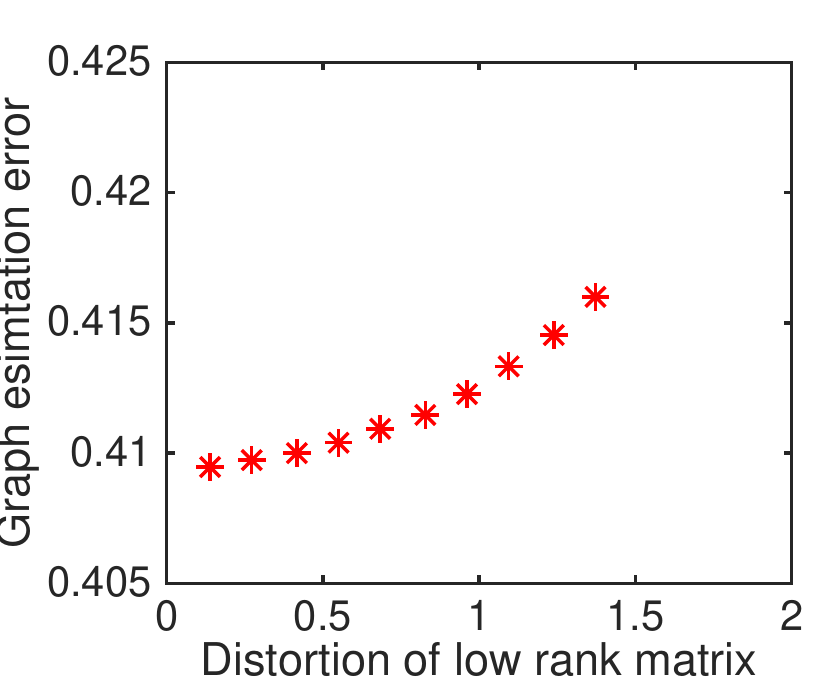}\tabularnewline
(a) & (b) \tabularnewline
\end{tabular}
\par\end{centering}
\caption{Graph estimation errors with different noisy low-rank components as inputs. In (a), the low-rank components have fixed noise amplitude but different distortion sparsity (i.e., $u$). In (b), the low-rank components have a fixed distortion sparsity ($u = 0.2$) but different distortion  amplitude (i.e., $c$).}
\label{fig:Step2Result}
\end{figure}

For the first distortion, we change $u$ from 0.05 to 0.1. For the second distortion, we set $u=0.2$ and change $c$ from 0.1 to 1.
Figure \ref{fig:Step2Result} depicts the results. The horizontal axis represent low-rank components matrix error $\|\hat{\textbf{L}} -\textbf{L}_0\|_F/\|\textbf{L}_0\|_F$ resulted from certain $u$ or $c$.
We use the mean value of five runs as the result for each setting.
The results suggest that large low-rank component distortions degrade the  graph estimation.

These experiment results further justify 
our integrated scheme for simultaneous low rank estimation and graph refinement for  signals that are smooth on graph.
In our LGE, the first step estimates the low-rank components from the noisy data with the help of the (inexact) graph. The second step refines the graph based on the estimated low-rank components. These two steps iterate and enhance each other until optimal solution has been obtained.

\subsection{Experiments on the entire LGE}
\label{ssec:DataNoiseSparsity}
After examining individual steps of LGE, in this section, we experiment the two steps of LGE together as a whole.



For perturbation on the input data, the assumption of LGE is similar to the RPCA\cite{Cand11}.
In particular,  
the perturbation matrix $\M$ is assumed to be sparse.
In~\cite{Cand11}, their synthetic experiment uses perturbation sparsity $d=0.05$ to $d=0.1$ (recall that
$d= \|M\|_0 / pn$). In this experiment, we further experiment the situations when $\M$ is less sparse. 
Specifically, we experiment with perturbation matrix $\M$ with $ d = 0.1$ to $ d = 1$ and with noise amplitude $\pm1$. We also experiment RPCA with this noisy data.
Low-rank components estimation errors are measured using $\|\hat{\textbf{L}} -\textbf{L}_0\|_F/\|\textbf{L}_0\|_F$.
In this experiment, we use two types of initial input graph.
One is the ground-truth graph $\Phi_{f0}$. The other is an approximated graph calculated from the data using k-nearest neighbor method. 
Specifically, for each node, it only connects to its 5 nearest neighbors based on the rows of input data $\X$. 



The results are listed in Table \ref{tab:SynResult_NoiseSparsity_gsp_Lap}. 
As shown in the table, when the perturbation sparsity is small, RPCA estimates the low-rank components better. However, as the perturbation sparsity in $\M$ increases (sparsity of perturbation $d\geq0.3$), LGE performs better. In LGE, the graph Laplacian help regularize the problem when perturbation sparsity is not small.  Even when the input graph is inexact (approximated using k-nearest neighbor method), the graph is refined iteratively in LGE and this leads to comparable gain as using the ground-truth graph, as shown in Table~\ref{tab:SynResult_NoiseSparsity_gsp_Lap}.

\begin{table}\small
	\caption{Low-rank components estimation error with different perturbation sparsity.}
	\label{tab:SynResult_NoiseSparsity_gsp_Lap}
	\centering
	\begin{tabular}{|*{1}{p{25pt}|}*{3}{c|}}
		\hline
		$d$ & LGE with $\Phi_{f0}$ & LGE with  $knn$ graph & RPCA\\
		\hline
		0.1 &  0.1339  & 0.0735  &\textbf{0.0001} \\
		\hline
		0.2 &  0.4171  & 0.3056  &\textbf{0.0722}\\
		\hline
		\textit{0.3} & \textbf{0.7139} & \textbf{0.6953} &  0.8666\\
		\hline
		\textit{0.4} & \textbf{0.8959}  & \textbf{0.8875} & 2.6751\\
		\hline
		\textit{0.5} &  \textbf{0.9617} & \textbf{0.9550} &  3.6462\\		
		\hline
		\textit{0.6} &  \textbf{0.9933}  & \textbf{0.9909}  & 3.9410 \\	
		\hline
		\textit{0.7} &  \textbf{0.9992}  & \textbf{1.0001} & 4.0338\\		
		\hline
		\textit{0.8} & \textbf{0.9228}  & \textbf{0.9205} & 3.4320\\		
		\hline
		\textit{0.9} &  \textbf{0.9978}  & \textbf{0.9996} & 4.1152\\		
		\hline
		\textit{1.0} &  \textbf{1.0113}  & \textbf{1.0305} & 4.1486 \\
		\hline
	\end{tabular}
	\vspace{+0.1cm}
	\normalsize
\end{table}

\subsection{Comparison with state-of-the-art}
\label{ssec:ComparisionDiffMethod}

In this experiment, we compare LGE with several state-of-art methods, including GL-SigRep\cite{Frossard:2016}, SGDict\cite{petric2017graph}, RPCAG\cite{Shah15a} and RPCA\cite{Cand11}.
GL-SigRep is a graph learning algorithm with some estimation of noise-free version of input data as by-product, based on a Gaussian noise model. 
SGDict is also a graph learning algorithm and it assumes that the graph signal is a sparse linear combination of a few atoms of a structured graph dictionary.
RPCAG is an improvement of RPCA using spectral graph regularization. It assumes the input graph is exact.
We compare these methods using the low-rank components estimation error: $\| \hat{\textbf{L}} -\textbf{L}_0\|_F/\|\textbf{L}_0\|_F$ and graph estimation error: $\| \hat{\Phi_f} -\Phi_{f0}\|_F/\|\Phi_{f0}\|_F$.  
Results are listed in Table \ref{tab:SynResult}. Note that RPCAG, RPCA only estimate low-rank components, while SGDict only estimates the graph.
For low-rank components estimation, LGE achieves the smallest estimation error. For graph estimation, LGE has a smaller estimation error than GL-SigRep, but a larger error comparing with SGDict. However, it should be noted that SGDict does not  estimate low-rank components. These results demonstrate that our proposed LGE is competitive in estimating low-rank components and the underlying graph simultaneously from non-Gaussian grossly-corrupted dataset.




\begin{table}\small
	\caption{Comparison of low-rank components estimation error and graph estimation error for synthetic data.}
	\label{tab:SynResult}
	\begin{tabular}{|*{1}{p{70pt}|}*{2}{c|}}
		\hline
		Methods&low-rank estimation&graph estimation\\
		\hline
		LGE & \textbf{0.0731} & 0.5112 \\
		\hline
		GL-SigRep&1.4554& 0.9850 \\
		\hline
		SGDict& - & \textbf{0.0881}\\
		\hline
		RPCAG&0.3825&-\\
		\hline
		RPCA&0.8666&-\\		
		\hline
	\end{tabular}
	\centering
	\vspace{+0.1cm}
	\normalsize
\end{table}

\section{Experiment with brain imaging data }
\label{sec:BrainData}
In this section, we apply our proposed LGE on a high-dimensional brain imaging dataset and compare it with other methods. 
We examine the performance of LGE in estimating  the low-rank components and brain connectivity graph from the high-dimensional data. This is practically useful for brain imaging studies: due to high dimensionality of data, low signal-to-noise ratio, and small number of available samples, it is challenging to estimate the low-rank approximation. 

The brain imaging dataset used here is a set of magnetoencephalography (MEG) signal
recorded at the MIT
Department of Brain and Cognitive Sciences \cite{hossein2016}.
A recumbent Elekta MEG scanner with 306 sensors was used to record the brain activity for 1100 milliseconds (100 milliseconds before and 1000 milliseconds after the presentation) for each stimulus. 
The two categories of visual stimuli were: 320 face images and 192 non-face images. These images were randomly selected and displayed passively with no task, and 16 subjects were asked to simply fixed at the center of the screen. All images were normalized for size and brightness among other factors, and were individually displayed once for 300 ms with random inter-stimulus delay intervals of 1000 ms to 1500 ms. 

We experiment LGE for these tasks: (1) estimation of low-rank components for classifying the signals evoked by face images or non-face images. (2) estimation of  the brain connectivity graph. 

\subsection{Initial graph}
\label{ssec:InitializationGraph}
In this experiment, we initialize $\Phi_f$ with the brain connectivity graph  generated from the \emph{resting state} measurements. The resting state in our experiment is 100ms of signal recording \emph{before} the stimuli presentation. Note that our method and all other methods are initialized with the same connectivity graph.

Three different types of brain connectivity graphs are commonly used in the literature: structural connectivity, functional connectivity and effective connectivity. Structural connectivity indicates the anatomical structure in the brain; functional connectivity quantifies functional dependencies between different brain regions; and effective connectivity indicates directed or causal relationship between different brain regions~\cite{bullmore2009complex}. 
After some evaluation, we found that we can achieve good accuracy with functional connectivity, specifically, {\em coherence connectivity}. 



Coherence connectivity quantifies oscillatory interdependency between different brain regions. It is a frequency domain analog of the cross-correlation coefficient. Given two series of signals $a_t$, $b_t$ for two brain regions $A$ and $B$ with a frequency $f$, first, decompose the signal spectrally at target $f$ to obtain the instantaneous phase at each time point~\cite{schmidt2014whole}. Then, filter each signal between $f\pm5Hz$ and calculate convolution of $f(t)$ with a Morlet wavelet centered at frequency $f$ to obtain the instantaneous phase at time $t$. Finally, the two signals can be represented as: $a=E_a(t)e^{j\psi_a(t)}$ and $b=E_b(t)e^{j\psi_b(t)}$, where $E_a(t)$ and $E_b(t)$ are amplitudes, $\psi_a(t)$ and $\psi_b(t)$ are the phases for $A$ and $B$ at time t. The coherence connectivity edge is computed as:

\vspace*{-0.5\baselineskip}
\begin{equation}\label{coh}
w_{A,B}=\left|\frac{\frac{1}{T}\sum_{t=1}^{T}E_a(t)E_b(t)e^{j[\psi_a(t)-\psi_b(t)]}}{\sqrt{\frac{1}{T}\sum_{t=1}^{T}E_a(t)^2}\cdot\sqrt{\frac{1}{T}\sum_{t=1}^{T}E_b(t)^2}}\right|
\end{equation}

After we obtain the adjacency matrix $\W$ with the above procedure, we can calculate the Laplacian matrix $\Phi_f = \D - \W$ as the initialization, where $\D$ is the diagonal degree matrix of $\W$.



\subsection{Experiment results with brain imaging data}
\label{ssec:BrainResult}

We apply LGE and related methods to estimate the low-rank components and apply a support vector machine (SVM) on the low-rank components
to classify whether the subject was viewing face or non-face images based on the MEG data. To evaluate low-rank components estimation accuracy, we compare these methods based on their classification accuracy. To evaluate graph estimation, we compare the compatibility of their connectivity graph matrix to the related neuroscience findings on cortical regions involving face processing.

MEG and EEG components corresponding to the face/non-face distinction have been reported at latencies of approximately 100 ms, and more reliably at 170 ms (also known as N170 marker, reported at about 145ms in MEG studies), after visual stimulus onset (e.g. see \cite{Perrett1987, Thorpe1996, Liu2002, Desimone2006}). In this experiment, we therefore choose the data from two time slots, namely 96ms to 105ms and 141ms to 150ms after the stimuli presentation, to be able to compare our automated assessment with the related neuroscience literature. 

To measure the performance of low-rank components estimation, we decompose low-rank components estimation $\hat{\textbf{L}}$ using SVD,  select the basis vectors based on the rank of $\hat{\textbf{L}}$, project the original data onto these vectors to obtain low-dimensional representations, and perform classification on the low-dimensional representations to obtain classification accuracies. We compare LGE with GL-SigRep, RPCAG, RPCA and PCA in the experiments.

Table~\ref{tab:BrainResult}
depicts the classification results for different methods on the two time slots with coherence connectivity matrix as the initial graph. \textit{P-value} for statistical hypothesis test is provided. In both time slots, the proposed LGE achieves the highest classification accuracy. Moreover, the \textit{P-value} of GL-SigRep, RPCAG, RPCA and PCA are less than 0.05, which is a very strong evidence against the null hypothesis. It indicates that LGE out-performs other methods for this task.

\begin{table}\small 
	\caption{Classification performance (accuracies) for brain imaging data (Time: 96ms-105ms and 141ms-150ms)}
	\label{tab:BrainResult}
	\begin{tabular}{|*{1}{p{45pt}|}*{4}{c|}}
		\hline
		&\multicolumn{2}{c|}{Time: 96ms-105ms}&\multicolumn{2}{c|}{Time: 141ms-150ms}\\
		\cline{2-5}
		Methods& SVM& P-value & SVM & P-value\\
		\hline
		LGE & \textbf{64.26\%}  & - &\textbf{79.53\%}  & -\\
		\hline
		GL-SigRep& 61.92\% & 2.32$\times 10^{-4}$ & 77.07\%  & 1.15$\times 10^{-5}$\\
		\hline
		RPCAG&62.55\% & 0.002 & 78.39\%  &  0.003\\
		\hline
		RPCA&59.92\% & 1.73$\times 10^{-6}$ &74.41\% & 2.23$\times 10^{-8}$\\		
		\hline
		PCA& 61.06\% & 2.14$\times 10^{-5}$ &75.44\%  & 1.89$\times 10^{-7}$\\
		\hline
	\end{tabular}
	\centering
	\vspace{+0.1cm}
	\normalsize
\end{table}

\begin{figure}\small
\begin{centering}
\begin{tabular}{cc}
\multicolumn{2}{c}{\includegraphics[width=0.40\columnwidth]{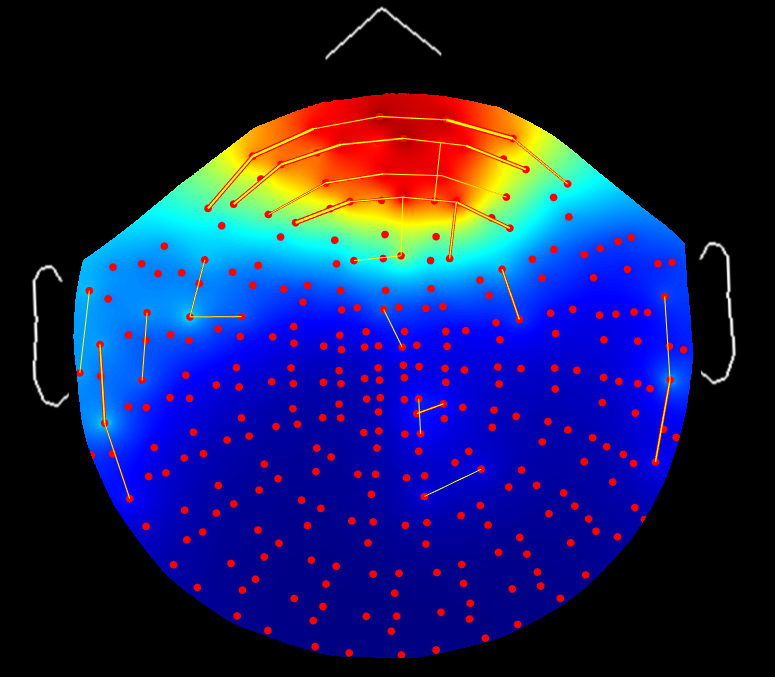}}\tabularnewline
\multicolumn{2}{c}{Initial graph: coherence connectivity at 50Hz}\tabularnewline
\includegraphics[width=0.48\columnwidth]{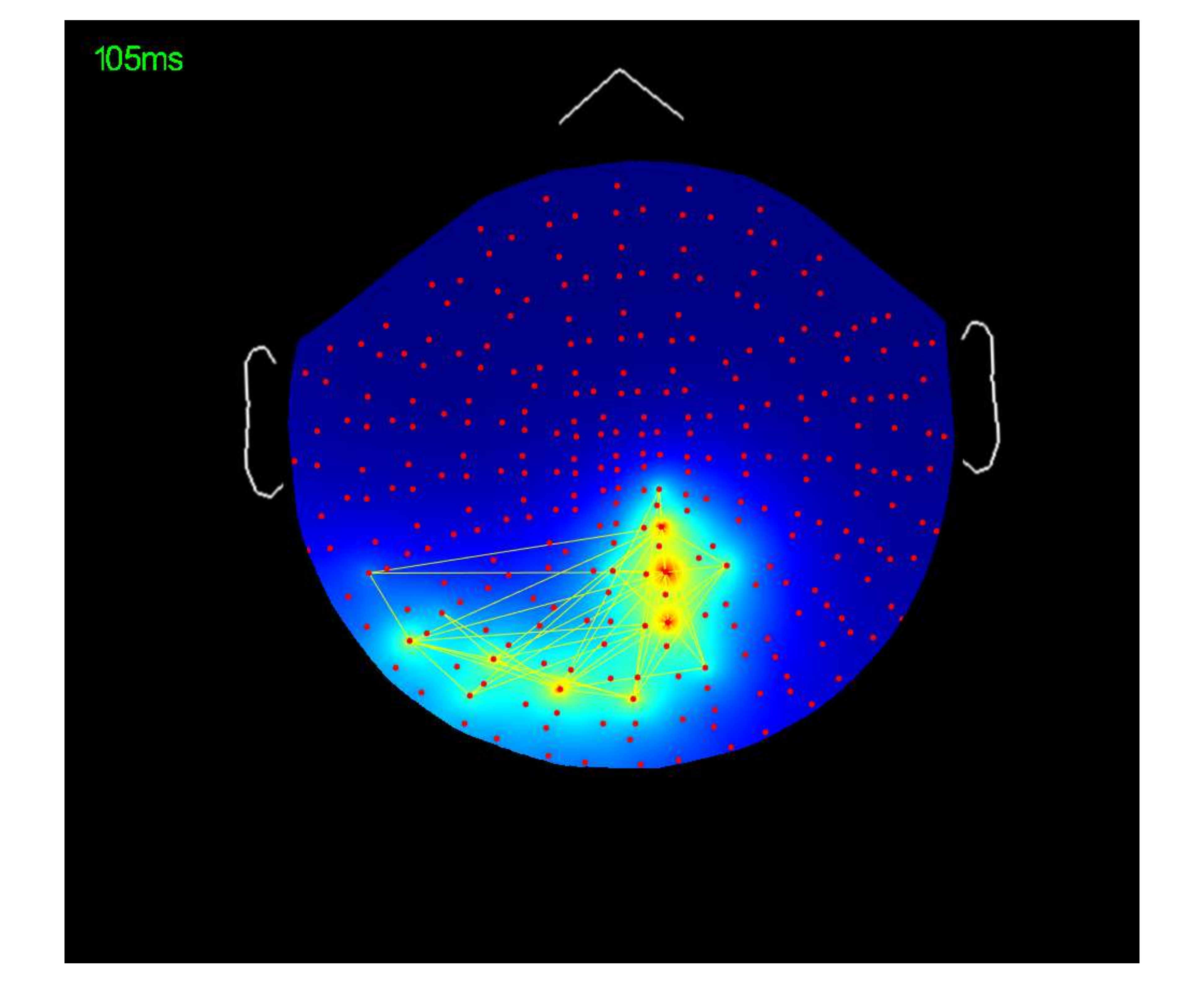} & 
\includegraphics[width=0.48\columnwidth]{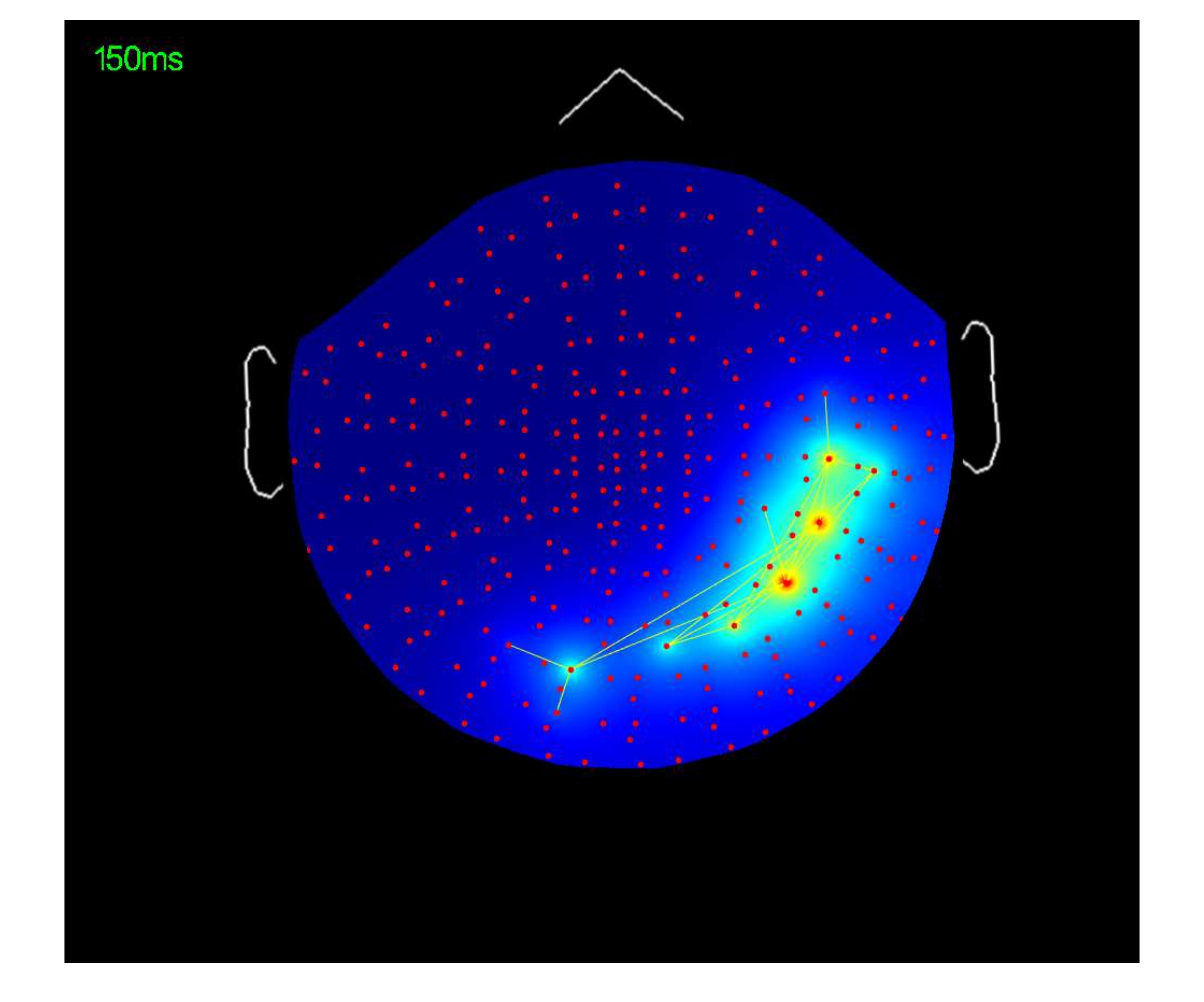}\tabularnewline
(a) LGE at 105ms & (b) LGE at 150ms\tabularnewline
\includegraphics[width=0.48\columnwidth]{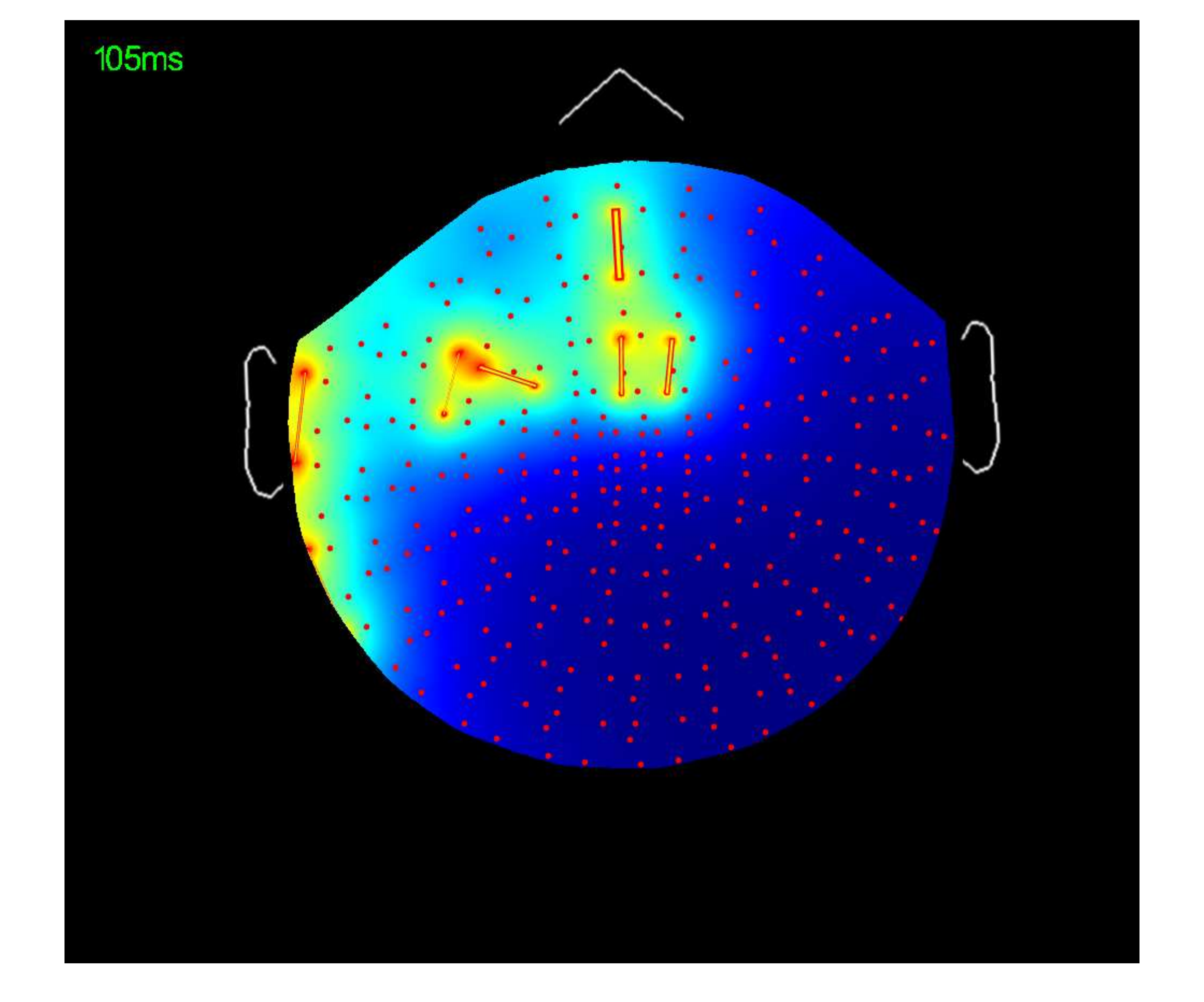} & 
\includegraphics[width=0.48\columnwidth]{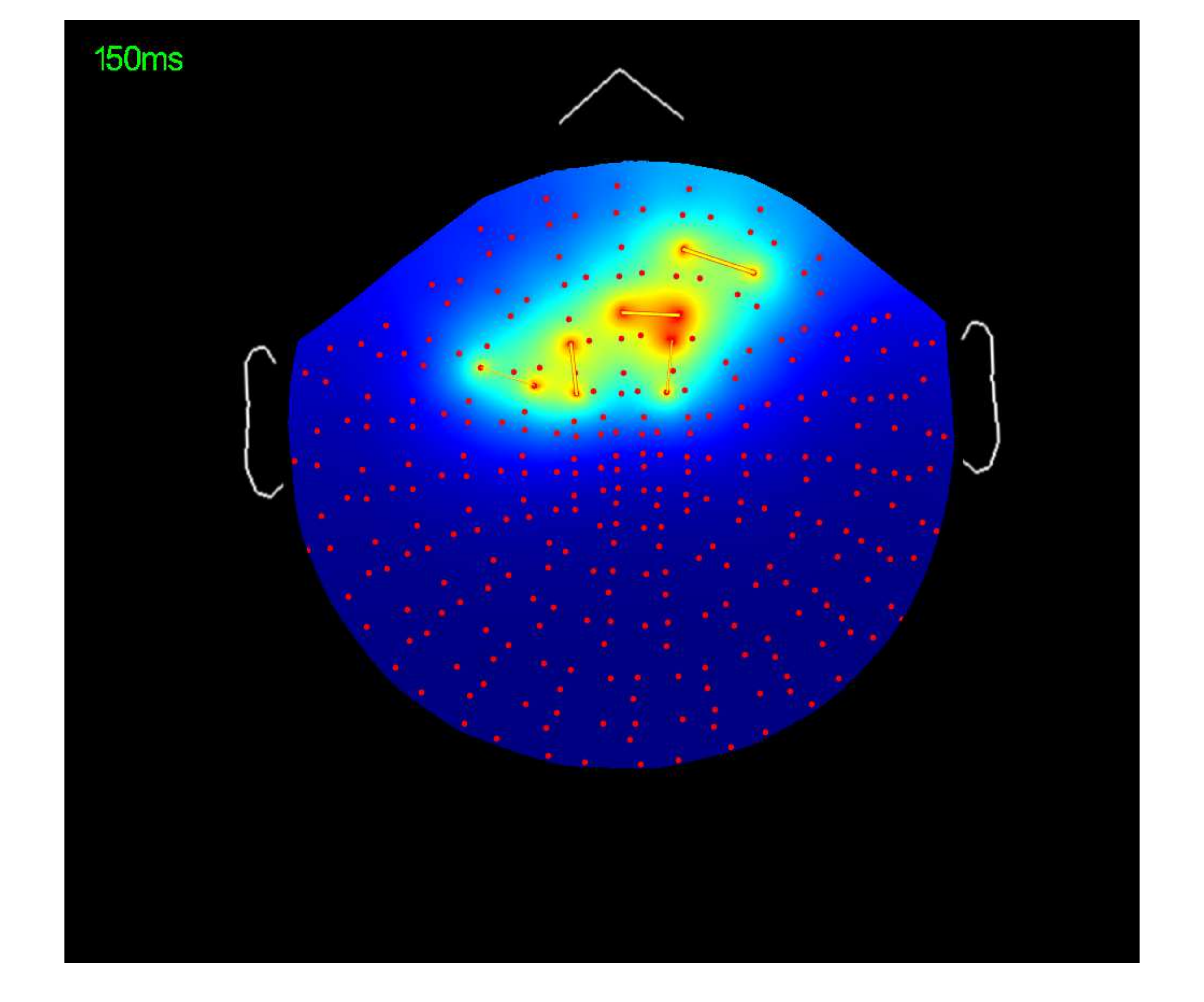}\tabularnewline
(c) GL-SigRep at 105ms& (d) GL-SigRep at 150ms\tabularnewline
\includegraphics[width=0.48\columnwidth]{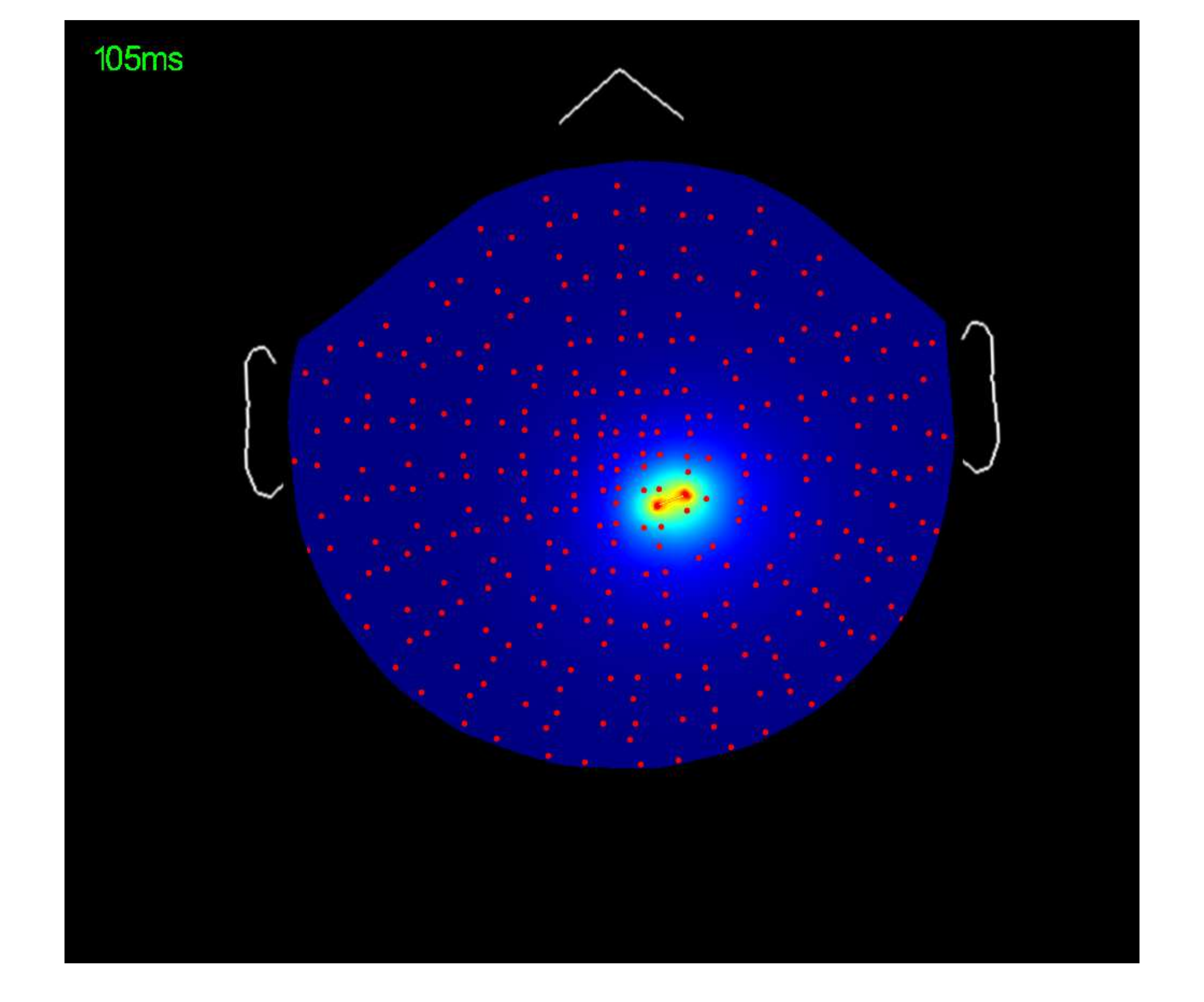} & 
\includegraphics[width=0.48\columnwidth]{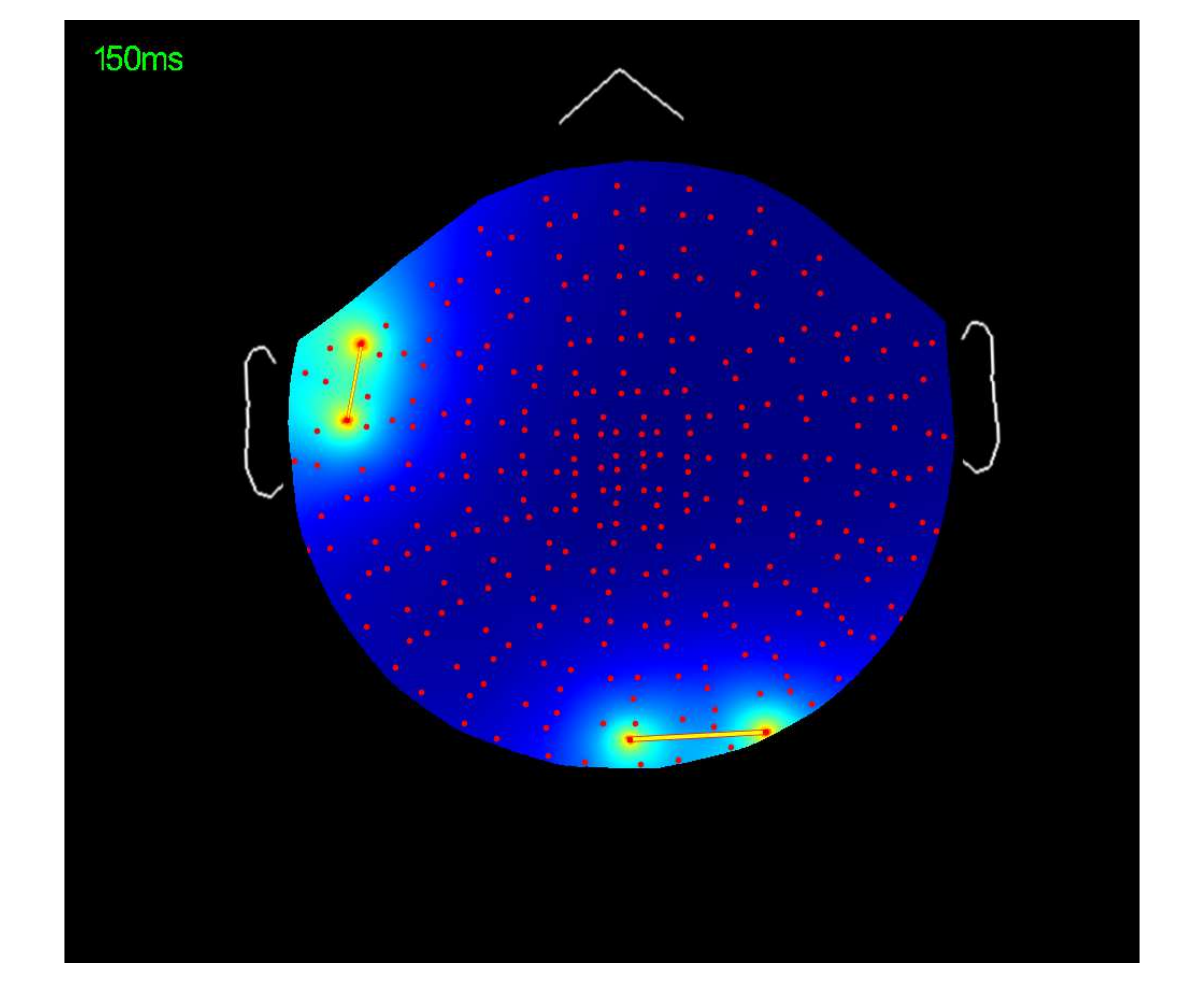}\tabularnewline
(e) SGDict at 105ms& (f) SGDict at 150ms\tabularnewline
\end{tabular}
\par\end{centering}
\caption{Graph estimation results for different graph learning algorithm.}
\label{fig:BrainFigure}
\end{figure}

To examine the results of graph estimation, we visualize the estimated graph Laplacian matrix $\hat{\Phi_f}$ by projecting the graph connectivity weights onto the MEG sensor locations.  We compare LGE with some state-of-art graph learning algorithm, including  GL-SigRep and SGDict. Figure \ref{fig:BrainFigure} depicts the results. The figure on the top row shows the initial graph
used for all methods, i.e., coherence connectivity at 50Hz estimated using the resting state measurements. The second row visualizes graph estimation of LGE at 105ms and 150ms. The third row visualizes the output of GL-SigRep at the same time instants. And the bottom row depicts the corresponding results of SGDict.

At 105ms after the presentation of visual stimuli (e.g., face images), neuroscientific literature reports that the visual information is still being processed at early visual cortex at occipital and occipitotemporal regions (e.g. see \cite{Thorpe1996,Desimone2006}). 
Examining the first column of 
Figure~\ref{fig:BrainFigure}, the graph connectivity estimation by LGE tends to span on the occipital and left occipitotemporal regions. Therefore, LGE seems to have reached a successful estimation of the true underlying connectivity at this time-point. However, GL-SigRep indicates connectivities at the left temporal and middle and inferior frontal gyri. SGDict indicates connectivities at the middle gyri. None of these regions have apparent link with early visual processing described in neuroscientific literature.  It should be noted that GL-SigRep assumes Gaussian noise and might not be appropriate for MEG data.

Examining the second column, the connectivity graph estimated by LGE at 150ms converges on connections on the right occipitotemporal region. This graph connectivity is comparable to the neuroscientific findings on face perception, specifically the N170 marker. In several studies such as \cite{Perrett1987,Liu2002}, the fusiform gyrus (at the occipitotemporal region) are suggested for processing face perception during about 145ms after presentation of face image stimuli, also known as N170 marker (named after its first discovery at 170ms in EEG studies). However, the estimated graph by GL-SigRep does not change significantly at 150ms. SGDict exhibits connectivities at occipital and left frontal region of the brain which indicates extra active region compared to neuroscientific findings. In summary, our technique reveals almost the same regions as important graph connections for face perception. The similarity of our estimated graph connectivity with neuroscientific literature supports that LGE is superior for this task.

\section{Conclusion}
\label{sec:conclusion}

Given a high-dimensional, graph-smooth dataset, we proposed an algorithm to learn the low-rank components and graph simultaneously and iteratively. We focused on the situations when the perturbations on the low-rank components are grossly but sparse. We analyzed
how the inexact graph impacts the low-rank components estimation.
We evaluated the proposed LGE using both synthetic and real-world data, i.e., brain imaging data. Comparison with other state-of-the-art methods suggests that LGE is competitive in both low-rank approximation and graph estimation. In addition, when applying to MEG data, LGE could recover connectivity graphs that are compatible to the neuroscientific literature.

\section*{Acknowledgment}

The authors would like to thank
Pavitra~Krishnaswamy of 
Institute for Infocomm Research, A*STAR 
for helpful discussion.

\ifCLASSOPTIONcaptionsoff
  \newpage
\fi



%

\bibliographystyle{IEEEtran}
\bibliography{IEEEabrv,refs}
\end{document}